\documentclass{pasj00}

\begin{document}

\SetRunningHead{Y.Shirasaki et al}{AGN and Galaxy Clustering at $z$ = 0.3 to 3.0}

\Received{2009/07/31}%{yyyy/mm/dd}
\Accepted{2010/10/30}%{yyyy/mm/dd}

\title{Early Science Result from the Japanese Virtual Observatory: 
AGN and Galaxy Clustering at $z$ = 0.3 to 3.0
\thanks{
Based on data collected at Subaru Telescope, which is operated by the National Astronomical Observatory of Japan (NAOJ).
Data is retrieved from the JVO (http://jvo.nao.ac.jp/portal) operated by NAOJ.}
}

%%% begin:list of authors
% Do NOT capitalize all letters in "textsc".
\author{
  Yuji \textsc{Shirasaki}\altaffilmark{1}
  Masahiro \textsc{Tanaka}\altaffilmark{2}
  Masatoshi \textsc{Ohishi}\altaffilmark{1}
  Yoshihiko \textsc{Mizumoto}\altaffilmark{1}
  Naoki \textsc{Yasuda}\altaffilmark{3}
  Tadafumi \textsc{Takata}\altaffilmark{1}
}

\altaffiltext{1}{National Astronomical Observatory of Japan, 
2-21-1 Osawa, Mitaka City, Tokyo, 181-8588, Japan}

\altaffiltext{2}{University of Tsukuba,
1-1-1 Tennodai, Tsukuba, Ibaraki, 305-8577, Japan}

\altaffiltext{3}{Institute for the Physics and Mathematics of the Universe Library,
The University of Tokyo,5-1-5 Kashiwa-no-ha, Kashiwa, Chiba, Japan}

\email{yuji.shirasaki@nao.ac.jp}

%%% end:list of authors

%%% `\KeyWords{}' always has to be placed before `\maketitle'.
 %Do NOT move this preamble from here!

\maketitle

\begin{abstract}
%
% less than 200 words 
%
We present the result of projected cross correlation analysis 
of AGNs and galaxies at redshifts from 0.3 to 3.0.
The Japanese Virtual Observatory (JVO) was used to obtain the
Subaru Suprime-Cam images and UKIDSS catalog data around AGNs.
We investigated 1,809 AGNs, which is about ten times larger 
a sample than that used in previous studies on AGN-galaxy clustering
at redshifts larger than 0.6.
90\% of the AGN samples are optically-selected AGN from the SDSS and
2dF catalogs.
The galaxy samples at low redshift includes many redder objects from 
UKIDSS survey, while at higher redshift they are mainly blue galaxies 
from Suprime-Cam.
We found significant excess of galaxies around the AGNs at redshifts 
from 0.3 to 1.8. 
For the low redshift samples ($z<0.9$), we obtained correlation length of
$r_{0} = $5--6~$h^{-1}$Mpc ($\gamma = 1.8$), which indicates that the AGNs
at this redshift range reside in a similar environment around typical 
local galaxies.
We also found that AGNs at higher redshift ranges
reside in a denser environment than lower redshift AGNs;
For $z=1.3 \sim 1.8$ AGNs, the cross correlation length was measured as
11$^{+6}_{-3}$~$h^{-1}$Mpc ($\gamma=1.8$).
Considering that our galaxies sample is based on optical observations
with Suprime-Cam at the redshift range, it is expected that blue star-forming 
galaxies comprise the majority of objects that are observed to be
clustered around the AGNs.
It is successfully demonstrated that the use of the archive through the 
Virtual Observatory system can provide a powerful tool for investigating the small 
scale environment of the intermediate redshift AGNs.

%
% For AGNs at the highest redshift samples ($z>1.8$), we could not obtain 
% a significant clustering signature, but derived an upper limit of 
% $r_{0} < 31$~h$^{-1}$Mpc.
%
% These results are compatible with a major merger scenario which incorporates
% downsizing effect of mass assembly.
%
% We also investigated dependence of the clustering property on 
% both the AGN's optical luminosity and redshift, and found that
% the correlation length for a brightest sample at redshift $z =$ 
% 1.0 -- 1.8 is twice as large as those obtained for lower luminosity
% samples.
%
%
% The high luminosity AGNs are produced at a contanct region of the two 
% colliding clusters or groups of galaxis.
%
% At higher redshift collisions between the larger systems are
% dominant, but it terminates rapidly and smaller scale collisions
% becomes dominant at lower redshift.
%
% The high luminosity AGNs transits to the low luminosty AGNs after
% a few Gyr.
%
% Our result implies that the Japanese Virtual Observatory can
% be a powerful tool to investigate the large scale structure 
% of the intermediate redshift Universe.
%
\end{abstract}

% \newpage

\section{Introduction}

It has been believed that the origin of AGN activity is accretion of matter
into a massive black hole at the center of a galaxy, and that almost all 
massive galaxies have such a black hole
(e.g. \cite{Salpeter_1964,Lynden-Bell_1969,Rees_1984,Kormendy_1995,Richstone_1998}).
In order to explain the activity of the luminous QSOs with black hole mass 
$\sim 10^{9}$ M$_{\Sol}$,
a large fraction of matter in the galaxy must be delivered to the inner region
of the galaxy on a short timescale, $\ll 10^{9}$~yr (\cite{Hopkins_2008}).
One possible mechanism for causing rapid gas inflows toward the central
region would be a major galaxy merger between gas-rich galaxies 
(e.g. \cite{Sanders_1988,Barnes_1991,Barnes_1996,Kauffmann_2000,Granato_2004,Springel_2005,Croton_2006}).
If this is the case, AGNs are expected to be found in an environment
\footnote{The term of environment is used to represent the excess number density
around the region relative to the average at the redshift.}
of higher density than typical local galaxies
\footnote{Models of merger rates suggest that the probability of major mergers increases
with denser environment until one reaches the small group scale of $M_{halo} \sim 10^{12}$ 
to $10^{13}$ M$_{\Sol}$/h, after which it decreases owing to higher velocity dispersions.
(e.g. see figure~4 of \cite{Hopkins_2008})
}.
There is some observational evidence that mass assembly 
of a larger galaxy terminates earlier than that of a smaller system 
(e.g. \cite{Cowie_1996, Kodama_2004}).
From these observational facts, it is expected that at higher redshift
the merger occurs in an environment with high galaxy density.
Thus the AGNs produced at the higher redshift should be observed in a
more crowded region.

Measurements of the two-point auto-correlation function of optically selected QSOs
were performed by using the large dataset of QSOs obtained by 2dF and SDSS surveys
(\cite{Croom_2005,Ross_2009,Shen_2009,Myers_2006,Myers_2007}).
\citet{Ross_2009} measured the real-space auto-correlation lengths for $z=$0.3--2.2
QSO, which was in the range of 6--11~$h^{-1}$Mpc for fixed power index of
$\gamma = 2.0$.
\citet{Shen_2009} analyzed data of QSOs at higher redshifts and obtained 
the real-space auto-correlation lengths of 14.58$\pm$2.70~$h^{-1}$Mpc ($\gamma=2.0$) and 
21.04$\pm$3.39~$h^{-1}$Mpc ($\gamma=2.0$) at $z=$2.9--3.5 and $z=$3.5--5.0, respectively.
The auto-correlation length of the local quiescent galaxy has been measured by many
authors (e.g. \cite{Brown_2000,Norberg_2002,Budavari_2003,Madgwick_2003,Hawkins_2003,Zehavi_2005,Ma_2009}).
%%
%% Madgwick_2003 --> 5.5+-0.3 h-1Mpc
%% Zehavi_2005 --> 5 h-1 Mpc
%% 
\citet{Zehavi_2005} measured real-space auto-correlation length and obtained
5.59$\pm$0.11~$h^{-1}$Mpc for flux-limited sample with $14.5 < r <17.77$ 
and $-22 < M_{r} <-19$ (for h = 1).
Thus the large scale environment of QSOs is almost equivalent to or only slightly 
larger than the environment of typical local galaxies at redshift less than 2.2, 
while at higher redshift QSOs are found in denser environments.
The direct measurement of QSO two-point auto-correlation function requires a large 
number of QSO samples and large sky coverage, thus only the large survey projects 
such as 2dF and SDSS can provide a precise and reliable measurement.
The measurement of QSO-galaxy cross correlation function, on the other hand,
provides a direct insight on the local environment of QSO with fewer samples, 
however a full 3-D QSO-galaxy cross correlation is limited to lower redshifts
where spectroscopic observation of galaxies is feasible.

A number of studies have been made to investigate the relation of QSO/AGN 
environment with a type and luminosity of the AGNs and their evolution with redshift
(e.g. \cite{Miller_2003,Sorrentino_2006,Coldwell_2006,Serber_2006,Strand_2008,Barr_2003,Coil_2007,Adelberger_2005,Bornancini_2007}).
%%
%%
%% SDSS にもとづく結果
%%
%%
Studies of the environment around optically selected AGNs at redshift $<0.6$ 
have been carried out by many authors by using the SDSS data.
\citet{Miller_2003} estimated the fraction of galaxies associated with an AGN, 
based on 4,921 galaxies with redshift less than $\sim$ 0.1.  
They found that the fraction of galaxies associated with an AGN is independent 
of density of local galaxy.
\citet{Sorrentino_2006} examined 1,829 AGNs at redshift $<0.1$, 
and found no evidence of a relation between the density of local galaxy
and the AGN activity.
\citet{Coldwell_2006} analyzed the environment of $\sim$2,000 AGNs at redshift 
$<0.2$, and found that the number density of galaxies around the AGNs is similar to 
that of a typical galaxy.

Although these results show that the optically selected AGNs do not 
reside in an area of high over-density, there are several studies
which indicate the small-scale over-density around the AGNs.
\citet{Serber_2006} analyzed small-scale environment of 2,028 $z < 0.4$ SDSS
QSOs and found that the QSOs are located in higher density region than are
$L^{*}$ galaxies and the over-density increases with decreasing scale at a distance 
less than 0.5~$h^{-1}$Mpc from the QSO. 
They also reported that there is a luminosity dependence of the density enhancement
at the small-scale and the brightest end.
\citet{Strand_2008} also investigated the relationship of the AGN environment 
to type (type I, type II), luminosity and redshift of the AGN at small
distance less than 2~$h^{-1}$Mpc.
They extended the redshift range up-to $0.6$ compared to the work of \citet{Serber_2006}
by applying a photometric redshift cut in their selection of galaxies to reduce the
projection effect.
The number of AGN samples was increased to $\sim 11,000$.
They concluded that the AGN environment depends on the luminosity of AGNs but not 
on the type of AGNs, and that there is marginal evidence on variation of 
its environment around type I QSOs as a function of redshift.
%%

%%
%% SDSS 以外
%%
Since the observation of galaxies by the SDSS is limited to the redshift of less than 
$\sim$0.6, measurements of AGN environment at higher redshifts need to be carried
out by an infrared instrument on a 2--4~m class telescope or optical
and/or near infrared observations by 8--10~m class telescopes.
\citet{Coil_2007} measured clustering around 52 SDSS and DEEP2 QSOs at 
redshifts 0.7 -- 1.4, by cross-correlating each QSO with DEEP2 galaxies.
They found that the clustering scale length between the QSOs and galaxies is comparable
to that obtained by auto-correlation of DEEP2 galaxies, and obtained the
cross-correlation length of 3.3$\pm$0.7~$h^{-1}$Mpc.
They found no significant dependence on luminosity nor redshift.
\citet{Adelberger_2005} examined clustering of galaxies around 79 AGNs with 
absolute magnitude from $-30$ to $-20$ at redshifts from 1.5 to 3.5, by cross 
correlating with the 1,627 galaxies with measured redshift.
The galaxy sample used in their analysis was constituted of
color-selected spectroscopic targets.
They divided 79 AGNs into two luminosity groups, and derived cross-correlation
lengths of 4.7$\pm$2.3~$h^{-1}$Mpc and 5.4$\pm$1.2~$h^{-1}$Mpc for the low 
and high luminosity groups, respectively.
They reported no evidence of luminosity dependence.
\citet{Bornancini_2007} examined the cross-correlation between the SDSS QSOs and 
distant red galaxies (DRG) at redshifts from 1 to 2, and obtained cross-correlation 
length of 5.4$\pm$1.6~$h^{-1}$Mpc.
All of these results for high redshift ($z=$0.7--3.5) AGN/QSOs also show
no significant over density of their environment.

Although the optically selected AGNs/QSOs seem to be located in an environment
comparable to that of local typical galaxies as introduced above, 
there have been some studies which report some types of QSOs such as radio-loud 
QSOs and X-ray selected AGNs are associated with a higher density region. 
\citet{Barr_2003} observed environments of 21 radio-loud QSOs at 0.6 $<$ $z$ $<$ 1.1
with multiple filters of optical to near infrared bands.
The observation showed that the radio-loud QSOs at this redshift range exist in a wide 
variety of environment (field galaxies, compact groups, and rich clusters), and 
indicated no evidence of redshift dependence of the environment.
They also found that the QSOs are not always located at the center of a galaxy 
group or cluster, which contrast with low redshift clusters where QSOs are located
at the center of galaxy distribution.
\citet{Enoki_2003} predicted that, at redshift from 1 to 2, QSOs are located
in various environments from small groups of galaxies to clusters of
galaxies based on a semi-analytic model of galaxy and QSO formation.
This prediction is consistent with the observational evidence obtained
by \citet{Barr_2003}.
The over density around the radio-loud QSOs were also reported by some 
other authors (e.g. \cite{Yee_1984,Yee_1987,Ellingson_1991,Hutchings_1999,Kauffmann_2008}).
In contrast, several studies reported that the environment of the
radio-loud and radio-quiet QSOs were indistinguishable 
(e.g.\cite{Fisher_1996,Wold_2001,McLure_2001}), though the number of samples used 
by these authors was not so large as several tens of samples.

There have been a number of studies on the environment of the X-ray selected AGNs 
(e.g. \cite{Mullis_2004,Grazian_2004,Gilli_2005,Yang_2006,Miyaji_2007,Hickox_2009,Gilli_2009,
Krumpe_2010,Coil_2009}).
\citet{Hickox_2009} examined the environment of AGN selected from radio
observations with Weterbork Synthesis Radio Telescope, X-ray observations
of Chandra XBootes Survey, and mid-IR observations of Spitzer IRAC Shallow 
Survey.
They measured cross-correlation between these AGNs and galaxies on scales of 
0.3--10~$h^{-1}$Mpc and derived the cross-correlation length of 6.3$\pm$0.6 
($\gamma = 1.8\pm0.2$),
4.7$\pm$0.3 ($\gamma = 1.6\pm0.1$), and 3.7$\pm$0.4~$h^{-1}$Mpc ($\gamma=1.5\pm 0.1$)
for AGNs selected in radio, X-ray, and infrared band.
They concluded that the X-ray selected AGNs were clustered similar to typical
galaxies, less clustered than radio AGNs, and more clustered than IR AGNs.
\citet{Krumpe_2010} examined the environment of X-ray selected AGNs, by using
cross correlation measurement between $\sim$ 1,550 ROSAT broad-line
AGN and $\sim$ 46,000 luminous red galaxies (LRGs) at $z = $0.16--0.36.
They obtained real space cross-correlation length of 6.93$^{+0.27}_{-0.28}$~$h^{-1}$Mpc
for the AGNs and LRGs, and also estimated the AGN auto-correlation length
of 4.3$^{+0.4}_{-0.5}$~$h^{-1}$Mpc.
They also measured an X-ray luminosity dependence of the clustering and 
concluded that low luminosity samples have a correlation length similar to
that of blue star-forming galaxies at low redshift, and high luminosity
samples have a larger correlation length than the lower luminosity samples
and is consistent with the clustering of red galaxies.

The auto-correlation of X-ray selected AGNs were measured by using data of
ROSAT all sky survey by \citet{Mullis_2004} and \citet{Grazian_2004},
Chandra Deep Field North and South by \citet{Gilli_2005} and \citet{Yang_2006},
XMM-COSMOS field by \citet{Gilli_2009} and \citet{Miyaji_2007},
and AEGIS field by \citet{Coil_2009}.
They obtained auto-correlation lengths of 6--10~$h^{-1}$Mpc for the X-ray
selected AGNs.
These results are slightly larger than
typical local galaxies and quite similar to the optically selected SDSS QSOs.
%%

%%
%% ここは書き直し
%%
Although the observations of AGN-galaxies clustering 
up to redshift $\sim$2.0 indicate that AGNs do not reside in a particularly 
high density environment, AGN auto-correlation study indicates that the 
environment of AGN at higher redshift is more crowded than in the low redshift
Universe.
Considering that the number density of bright AGNs peaks around redshift
$\sim$2 (\cite{Ueda_2003, Richards_2006}), it is expected that the evolution
of AGN is strongly affected by the surrounding galaxies at redshift $\sim$2.
It is, therefore, important to extend the study of AGN environment up to redshift 
two and beyond to solve the mechanism of co-evolution of AGNs and galaxies.
%%

%%
%% 本研究の目的・特徴など
%%
In this paper we present the result on measurements of AGN environment at redshifts 
from 0.3 to 3.0, which is the first scientific result obtained from the 
Japanese Virtual Observatory (JVO).
The reduced Subaru Suprime-Cam images are open to public through the JVO system.
The JVO portal provides a functionality to search for multiple regions
from its image data service by issuing a single JVOQL, which is an extension of 
database language SQL and can describe a coordinate join between the Suprime-Cam
image metadata table and a QSO/AGN catalog table.
The UKIDSS World DR2 is also used to study galaxies which are
dark in optical bands but bright enough to be detected in near-infrared bands.
The total number of QSO/AGN samples used in this work is $\sim$ 1,809, and $\sim$90\%
of the QSO/AGN samples comes from the SDSS and 2dF QSO catalog.
Note that this work uses the largest sample with deep optical images,
which are typically deeper than 24 mag.
By using the deep optical images, we can also measure the clustering 
of faint and blue star forming galaxies around AGNs at high redshift 
($z \sim$ 1 -- 2), which have not been well explored by any other studies.
Note also that our work would be free from the cosmic
variance, since the AGNs we have used are distributed through a wide area of the sky.

Throughout this paper, we assume a cosmology with $\Omega_{m} = 0.3$, $\Omega_{\lambda} = 0.7$
and $h = 0.7$.
All magnitudes are given in the AB system.
All the distances are measured in comoving coordinates.
The correlation length is presented in unit of $h^{-1}$Mpc.

% \clearpage

\section{The Datasets}
\label{sec:dataset}

\begin{figure*}
  \begin{center}
      \FigureFile(\textwidth,80mm){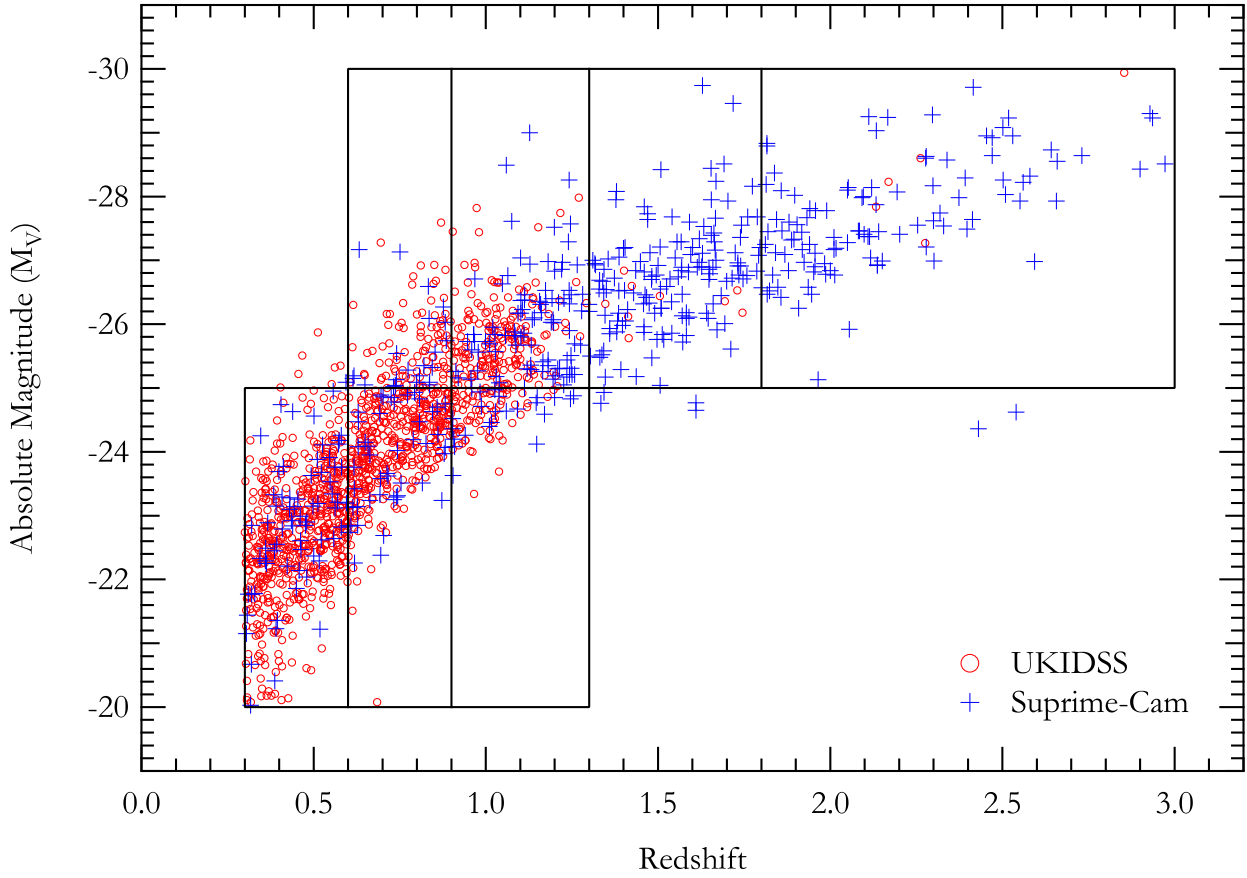}
  \end{center}
  \caption{K-corrected V band absolute magnitude vs redshift of the AGNs used in this work.
Open circles represent AGN samples for which the galaxy sample is derived from the UKIDSS data,
and the crosses represent AGN samples for which the galaxy sample is derived from the Suprime-Cam
data.}
  \label{fig:M_z}
\end{figure*}

\begin{figure*}
  \begin{center}
      \FigureFile(0.8\textwidth,80mm){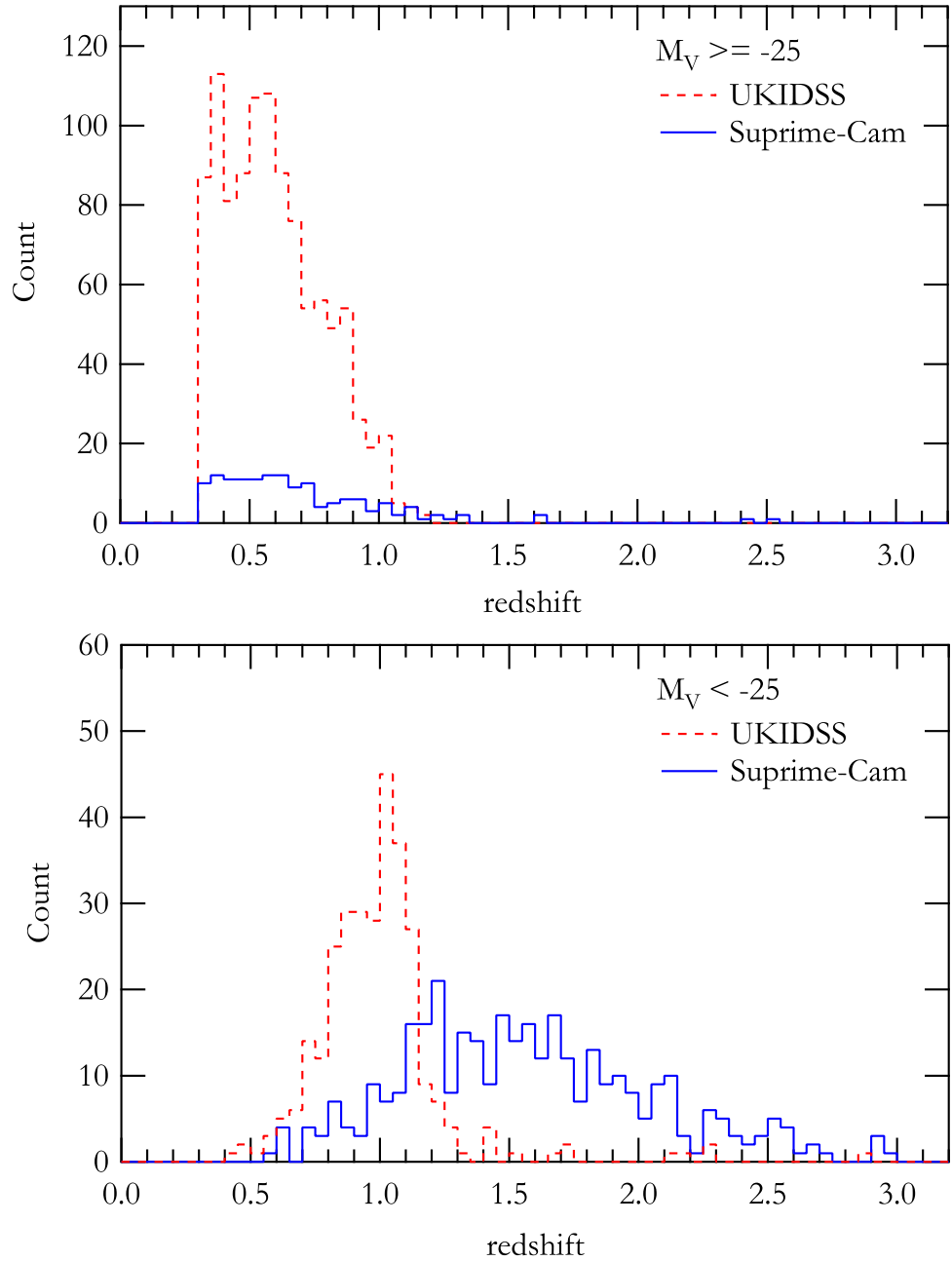}
  \end{center}
  \caption{Redshift distributions for dim AGN samples (top) and bright AGN samples (bottom).
The dashed histogram represents AGN samples for which the galaxy sample is derived from the UKIDSS
data, and the solid line histogram represents AGN samples for which the galaxy sample is derived
from the Suprime-Cam data.}
  \label{fig:zhist_D}
\end{figure*}

\begin{figure*}
  \begin{center}
      \FigureFile(\textwidth,80mm){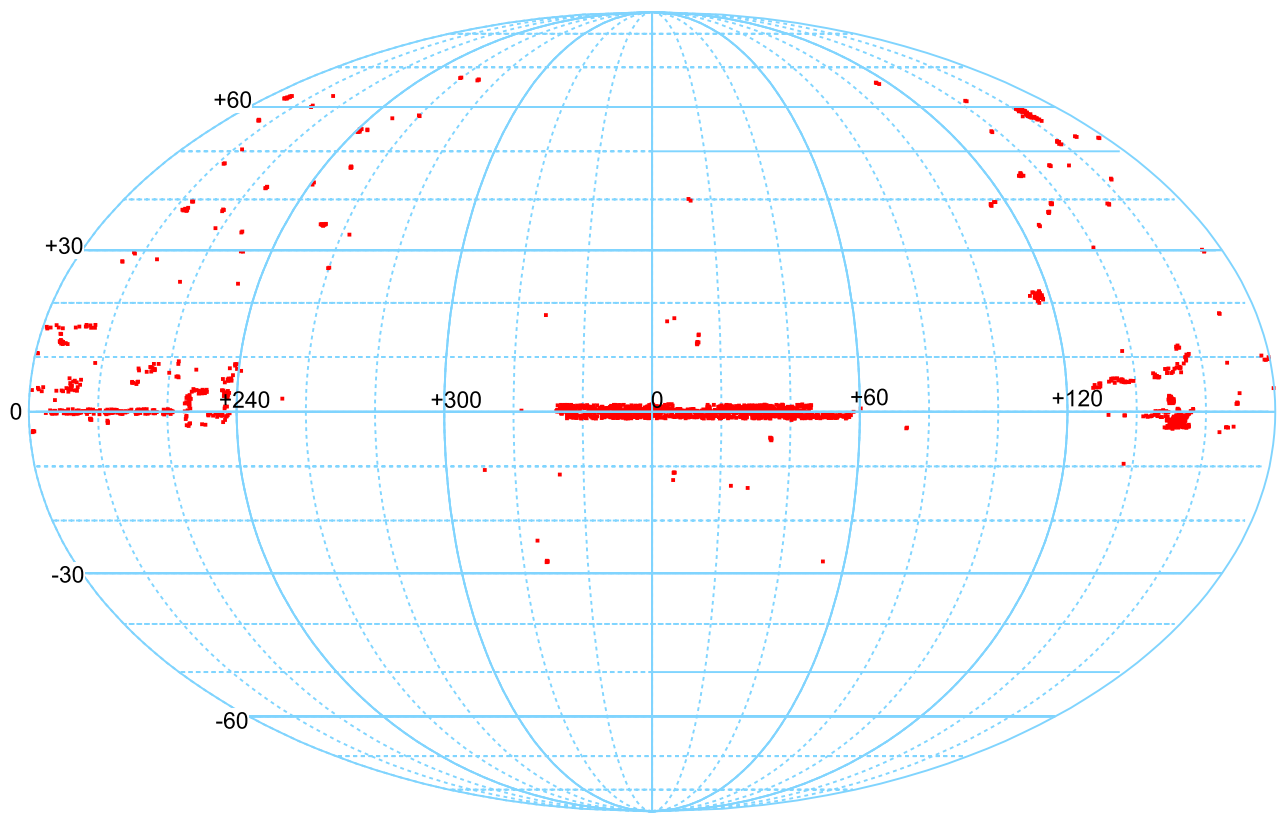}
  \end{center}
  \caption{The distribution of 1,809 AGNs analyzed in this work. 
           The coordinates are in equatorial.}
  \label{fig:skymap}
\end{figure*}

\begin{figure*}
  \begin{center}
      \FigureFile(0.9\textwidth,80mm){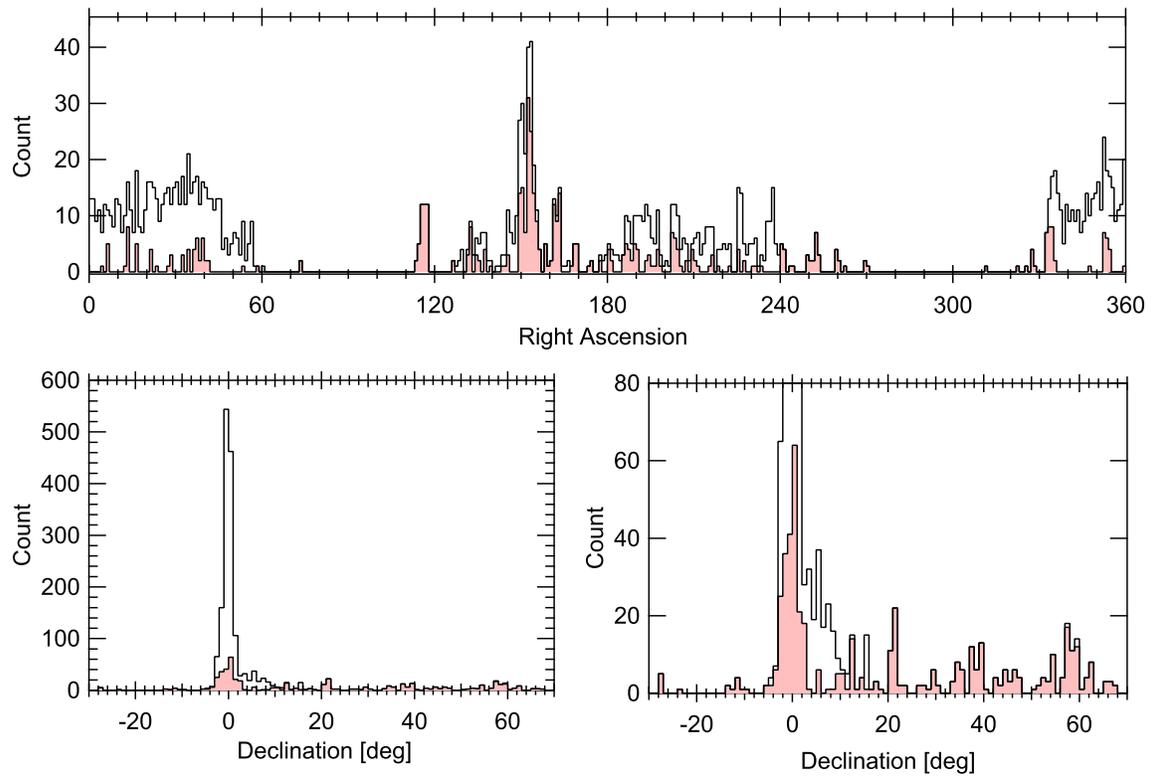}
  \end{center}
  \caption{The distribution of AGN coordinates. The top panel is for the right ascension 
of AGN,
the left bottom panel is for the declination, and the right bottom panel is a close up of 
the declination distribution, showing detailed distribution of low counts.
Open histogram is for all the AGNs  and the shaded histogram is for AGNs for which 
the Suprime-Cam data was used.}
  \label{fig:coord_hist}
\end{figure*}

All the data used in this work were obtained from the Japanese Virtual 
Observatory (JVO) portal (http://jvo.nao.ac.jp/portal).
The datasets downloaded through the JVO portal are:
Catalog of Quasars and Active Galactic Nuclei (12th Ed.) by \citet{Veron_2006},
SDSS DR-5 Quasar Catalog (4th Ed.) by \citet{Schneider_2007},
Subaru Suprime-Cam reduced images (version 0.1.24) of the JVO Subaru archive, 
and UKIDSS DR2 catalog by \citet{Warren_2007}.
90\% of the AGN samples used in this work are optically-selected AGN from the 
SDSS and 2dF catalogs.
The SDSS and 2dF QSOs are broad-line objects targeted on the basis of optical
photometry.
The distribution of absolute magnitude, redshift, and celestial coordinates of
the AGN samples are shown in figures~\ref{fig:M_z} to 
\ref{fig:coord_hist}.
Below we describe the procedure for selecting data in order to achieve our research
objectives.

%%
%% Suprime-Cam または UKIDSS データのある AGN の選択
%%
First we searched for AGNs observed with the Suprime-Cam or the 
UKIDSS survey at the redshift range of 0.3--3.0.
We selected AGNs whose K-corrected absolute $V$ band magnitude was 
in the range from $-30$ to $-20$.
%%
%% Quasar spectra and the K correction (2000A&A...353..861W)
%% ApJ 235 361 (1980) Richstone & Schmidt
%%  Schmidt & Green 1983   ApJ 269, 352
%%  Boyle et al 1988      MNRAS 235,935
%%
The K correction was made by assuming power law index of $\alpha=-0.5$
(e.g. \cite{Schmidt_1983,Boyle_1988})
for AGN spectrum ( $f_{\nu} \propto \nu^{\alpha}$ )
and calculated as $k = -2.5 (\alpha + 1) \log(1 + z)$.
According to \citet{Kennefick_2008} the variation of $\alpha$ 
was measured to be $\pm$0.4 for 45 QSO samples, 
which corresponds to error of one magnitude for k-correction at $z$ = 2.0. 
We don't expect that this error affects to our conclusion, since we did 
not observe any luminosity dependence as will be discussed in the later
section.
%%
%% 2Mpc フィルター
%%
Then we removed AGNs that were located near another lower redshift 
AGN within an angular distance corresponding to 8~Mpc at the AGNs'
redshift.
This filtering aimed to reduce the projection effect of galaxy 
clustering associated with other nearby AGNs.

%%
%% 一様度とデータカバレッジ 
%%
For each AGN among these samples, the Suprime-Cam images of B, V, Rc, Ic, i', and z' 
band and the UKIDSS K-band data were retrieved, and uniformity of the images and 
data coverage for the AGN field were examined.
%%
%% バックグラウンドの RMS
%%
The quality of some of the Suprime-Cam images retrieved from the JVO
is highly inhomogeneous, because the images were created by stacking all
the frames.
Certain parts of an image, therefore, may have significantly shorter exposure time
than other parts.
The inhomogeneity was estimated from the distribution of root mean squares (RMS)
of the background noise, which was calculated by the SExtractor (\cite{Bertin_1996})
at every image pixel.
We removed images where the standard deviation of RMS over the entire
AGN field was larger than 10\% of the average.
%%
%% $\sim$30\% of the images were removed with this criterion.
%%
The Suprime-Cam images which were not associated with photometric 
calibration data were not used in this analysis either.

%%
%% For each image, we masked regions around bright objects.
%%
Since we were interested especially in an environment of high redshift
($z>1$) AGN, we masked bright and extended sources on the Suprime-Cam images,
so that low redshift bright galaxies were effectively filtered from our galaxy 
sample.
The masked regions were defined as a region where
more than 100 contiguous pixels were illuminated with a flux higher than 10 
sigma of the background fluctuation.
Comparing with an unmasked image, we estimated that the masked sources were
about two magnitudes brighter than the peak of the magnitude distribution, 
which was $\sim$22 mag on average.
Object catalogs for Suprime-Cam images were created separately for each 
observation band using the SExtractor (\cite{Bertin_1996}).
%%
%% Here, we define an AGN field as a circular region of 2~Mpc radius centered
%% at the coordinate of an AGN.

%%
%% データカバレッジ
%%
In order to examine the data coverage of the AGN field,
an effective observation area was calculated for each annulus centered on an
AGN location, which has a corresponding width of 0.1~Mpc.
The effective area observed with the Suprime-Cam was calculated by counting the 
unmasked pixels of the image.
The coverage of the UKIDSS data was estimated from a fraction of an area
where number density of objects was lower than the average at the same
annulus by five sigma.
The density was estimated from a minimum azimuth angle between two objects
in the same annulus. 
The azimuth angle is an angular measurement in a spherical coordinate system 
with the origin at AGN location.
If the azimuth angle is larger than the average by five sigma, the region
between the two objects is considered to be an unobserved area.

It is difficult, however, to measure the coverage fraction of the annulus for 
which the expected source number is too small to measure the density variation
in the annulus.
For those annuli, we simply examined whether the average density is below
a threshold.
If it is below the threshold, the coverage fraction for the annulus was
considered to be 0\%, thus the sample was not used for the analysis.
If it is equal to or above the threshold, the coverage fraction estimated from
density variation, which is usually 100\% for the low count annulus, was used.
To determine the threshold, we estimated the average number
density expected for full coverage, and compared the observed number in the
annulus with the expected number.
The average number density was estimated from the annulus with second largest density
among the top-five annuli in terms of the number of detected sources.
If the observed number is below the expected number by five sigma, we considered
the annulus to be an unobserved area, that is 0\% of coverage.
We set a threshold on the data where at least 80\% of every annulus 
should be observed.
The data that did not meet this threshold were removed.
The fraction of data coverage estimated as described above was used in 
the calculation of number density of galaxy at each annulus.
After these selections, 2,689 AGNs were selected for further analysis.
%%

%%
%% 浅いデータの削除
%%
To maximize the signal to noise ratio, we removed the data of shallow 
observations based on the average number density, $\rho_{0}$, of detectable 
galaxies at the AGN redshift.
$\rho_{0}$ was estimated by integrating a luminosity function up to an effective
limiting magnitude, $m_{\mbox{\scriptsize limit}}$, of each Suprime-Cam image or UKIDSS
data.
The detailed description for determination of $\rho_{0}$ will be described in the
next section.

\begin{figure*}
  \begin{center}
      \FigureFile(\textwidth,80mm){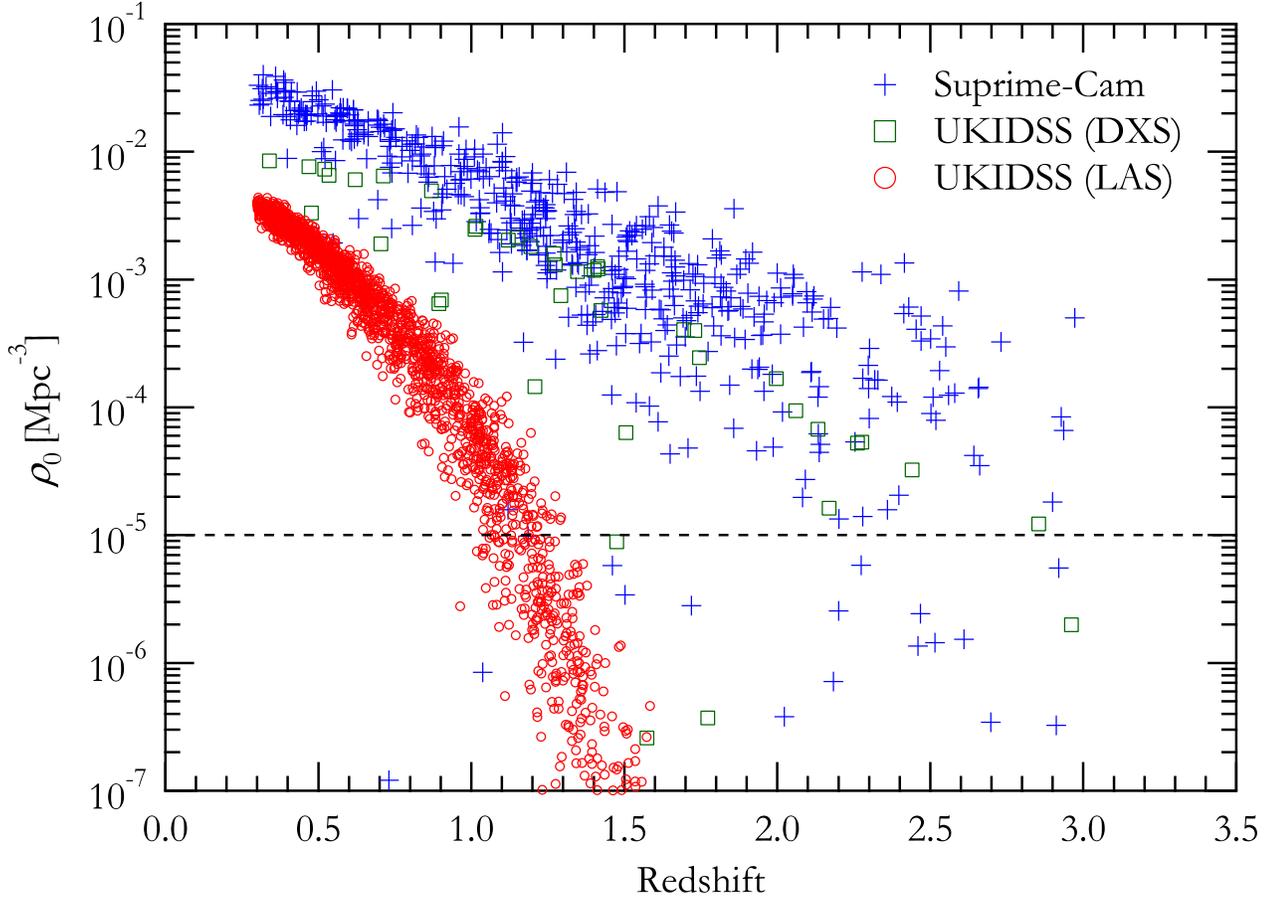}
  \end{center}
  \caption{The distribution of the average number density $\rho_{0}$ of
detectable galaxies at the AGN redshift $z$. Data of $\rho_{0} > 10^{-7}$ Mpc$^{-3}$
among the 2689 samples are plotted.
The open circles, boxes and crosses represent AGN samples for which the galaxy sample 
is derived from the UKIDSS Large Area Survey (LAS), Deep Extragalactic Survey (DXS),
and the Suprime-Cam archive data, respectively.
The horizontal dashed line represents the threshold for data selection.}
  \label{fig:rho0_z}
\end{figure*}

Then we selected data where $\rho_{0}$ was greater than $10^{-5}$~Mpc$^{-3}$ 
at any one of the observation bands.
The distribution of $\rho_{0}$ vs AGN redshift $z$ are shown in figure~\ref{fig:rho0_z}
for data of $\rho_{0} > 10^{-7}$.
In the figure, $\rho_{0}$ is the largest value among all the observation bands
for each AGN.
One can see a clear separation between the samples which are observed in
the UKIDSS Large Area Survey (LAS) (lower cluster of open circles), and the 
samples observed with the Suprime-Cam and UKIDSS Deep Extragalactic Survey (DXS)
(upper cluster of crosses and open circles).
This is because UKIDSS LAS data are significantly shallower than the data of
UKIDSS DXS and Suprime-Cam.
We applied the selection criterion of $\rho_{0} > 10^{-5}$ Mpc$^{-3}$,
which limit the maximum redshift of UKIDSS LAS samples to $\sim$1.3.
The number of AGN selected with this criterion is 2,023.
%%

\begin{figure*}
  \begin{center}
      \FigureFile(\textwidth,80mm){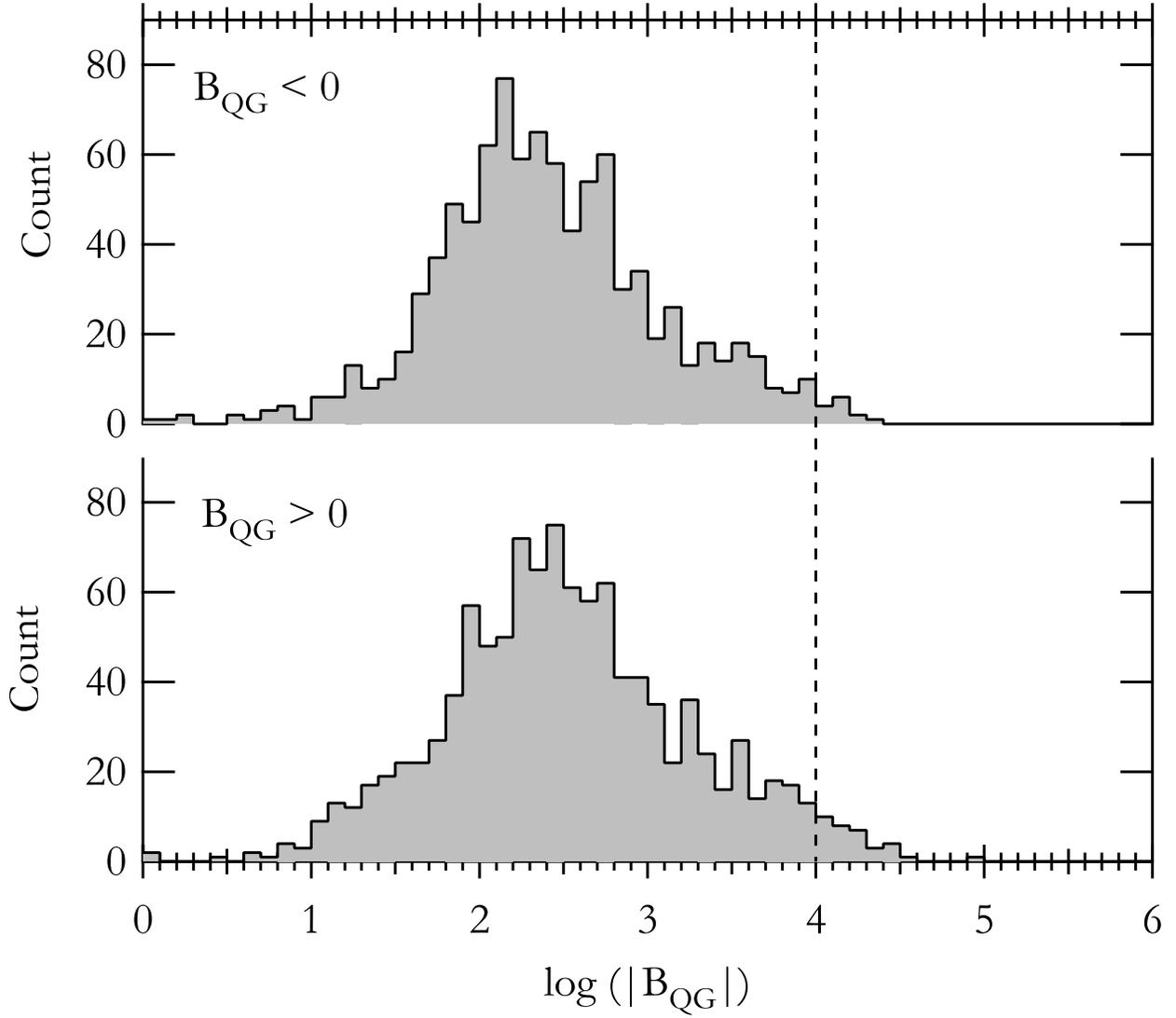}
  \end{center}
  \caption{The logarithmic distribution of clustering coefficient $B_{QG}$ for 2023 samples
which are selected with the criterion $\rho_{0} > 10^{-5}$ Mpc$^{-3}$.
The top panel is for negative $B_{QG}$ and the bottom panel is for positive $B_{QG}$.
The vertical dashed lines represent the upper limits for data selection.}
  \label{fig:B_QG_hist}
\end{figure*}

%% 
%% Bqg カット
%%
The foreground cluster or group of galaxies can be a noise for the clustering analysis.
To reduce the effect of accidental alignment of the nearby cluster, 
we calculated a clustering coefficient, $B_{\mbox{\scriptsize QG}}$, around each AGN, which was defined as
$\xi(r) = B_{\mbox{\scriptsize QG}} r^{-\gamma}$, and $B_{\mbox{\scriptsize QG}}$ was calculated as (\cite{Barr_2003}):
\begin{equation}
   B_{\mbox{\scriptsize QG}} = \frac{3-\gamma}{2\pi} 
   \frac{N_{\mbox{\scriptsize total}}- N_{\mbox{\scriptsize bg}}}{\rho_{0}}(1\mbox{Mpc})^{\gamma-3},
\end{equation}
where $N_{\mbox{\scriptsize total}}$ is the total number of observed galaxies at $r_{p} <$ 
1~Mpc, and $N_{\mbox{\scriptsize bg}}$ is the expected background count at $r_{p} <$ 1~Mpc estimated 
from the averaged density at $r_{\mbox{\scriptsize bg,min}} < r_{p} < r_{\mbox{\scriptsize bg,max}}$.
$\gamma$ was fixed to $1.8$.
The distance ranges $r_{\mbox{\scriptsize bg,min}}$ and $r_{\mbox{\scriptsize bg,min}}$ were determined
for each redshift group, and they are summarized in table~\ref{tab:distance_range_bg}.
If a cluster of galaxies is accidentally located in front of the AGN, 
significantly large $B_{\mbox{\scriptsize QG}}$ value would be observed.
Figure~\ref{fig:B_QG_hist} shows the distribution of $\log{|B_{\mbox{\scriptsize QG}}|}$ for 2,023 AGNs which
satisfied the $\rho_{0}$ criterion.
%%
%% $B_{qg}$ might become large due to the actual association of a galaxy cluster with
%% the AGN.
%%
%% We removed such extreme samples, as it significantly increases the average clustering 
%% strength.
%%
The selection criteria used here were $|B_{\mbox{\scriptsize QG}}| \le 1\times10^{4}$.
Abell class 0 clusters corresponds to $B_{\mbox{\scriptsize QG}} \sim$ 300 Mpc$^{1.77}$
(\cite{McLure_2001}).
The upper bound of this selection range corresponds to a cluster which has 30 times 
larger clustering coefficient than the Abell class 0, and it is expected that 
$B_{\mbox{\scriptsize QG}}$ of a real system rarely exceeds this criteria.
Thus most of the samples rejected with this criteria are the ones with an uncertainty 
of the clustering coefficient as large as this boundary or clusters
accidentally located near the line of sight.
The lower limit of $B_{\mbox{\scriptsize QG}}$ rejects a sample for which nearby cluster is accidentally 
located at the background region.
%%
%% The number of AGNs selected with this $B_{\mbox{\scriptsize QG}}$ criteria was 3570.
%%
Applying both the $\rho_{0}$ and $B_{\mbox{\scriptsize QG}}$ criteria, the number of AGNs was
reduced to 1,976.

\begin{figure*}
  \begin{center}
      \FigureFile(\textwidth,80mm){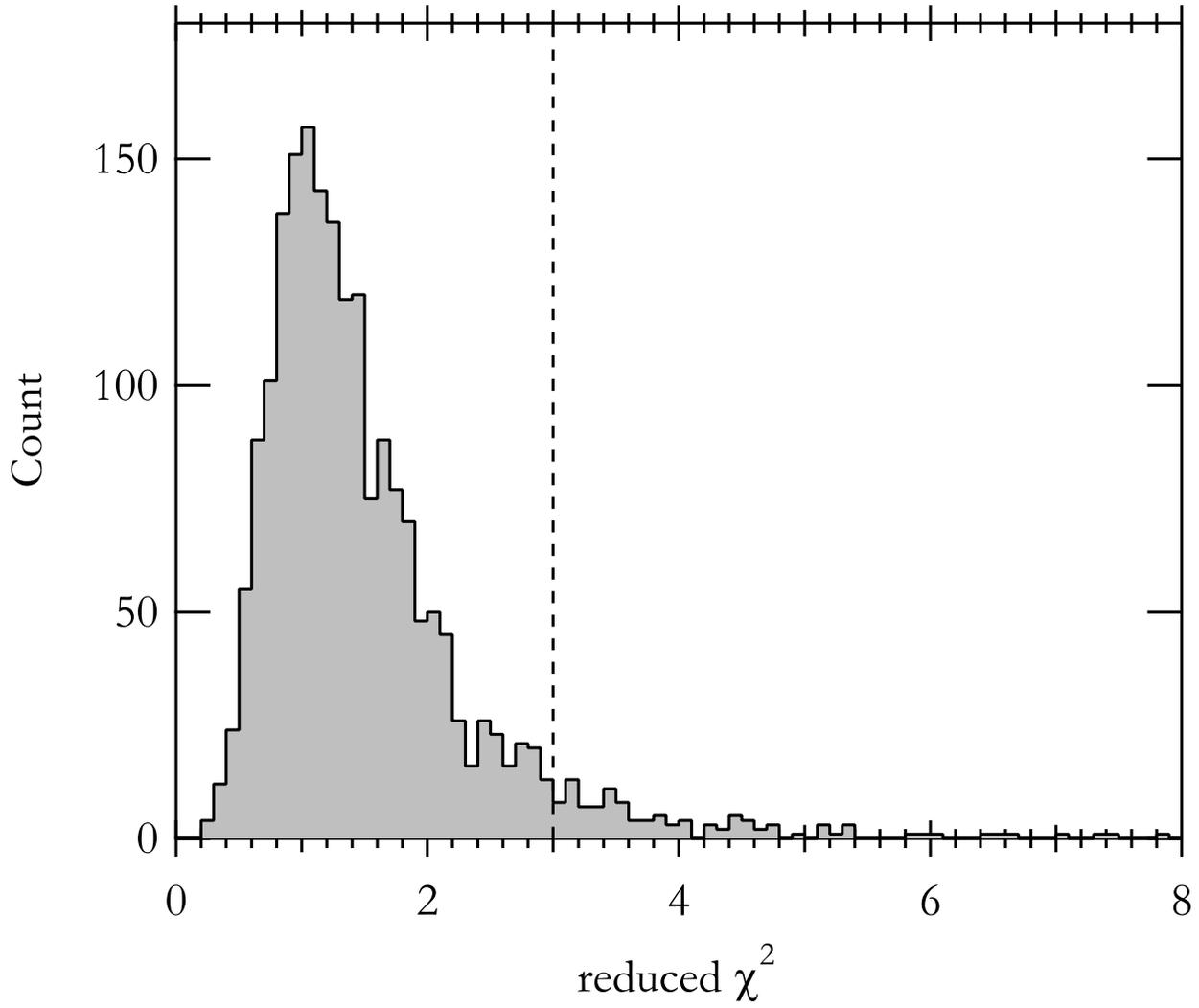}
  \end{center}
  \caption{The distribution of reduced $\chi^{2}$ of number density at 
radial distance from 1~Mpc to  $r_{\mbox{\scriptsize bg,min}}$ for 1976 AGN samples.
The vertical dashed line represent the maximum value for sample selection.}
  \label{fig:chi2}
\end{figure*}

\begin{figure*}
  \begin{center}
      \FigureFile(\textwidth,80mm){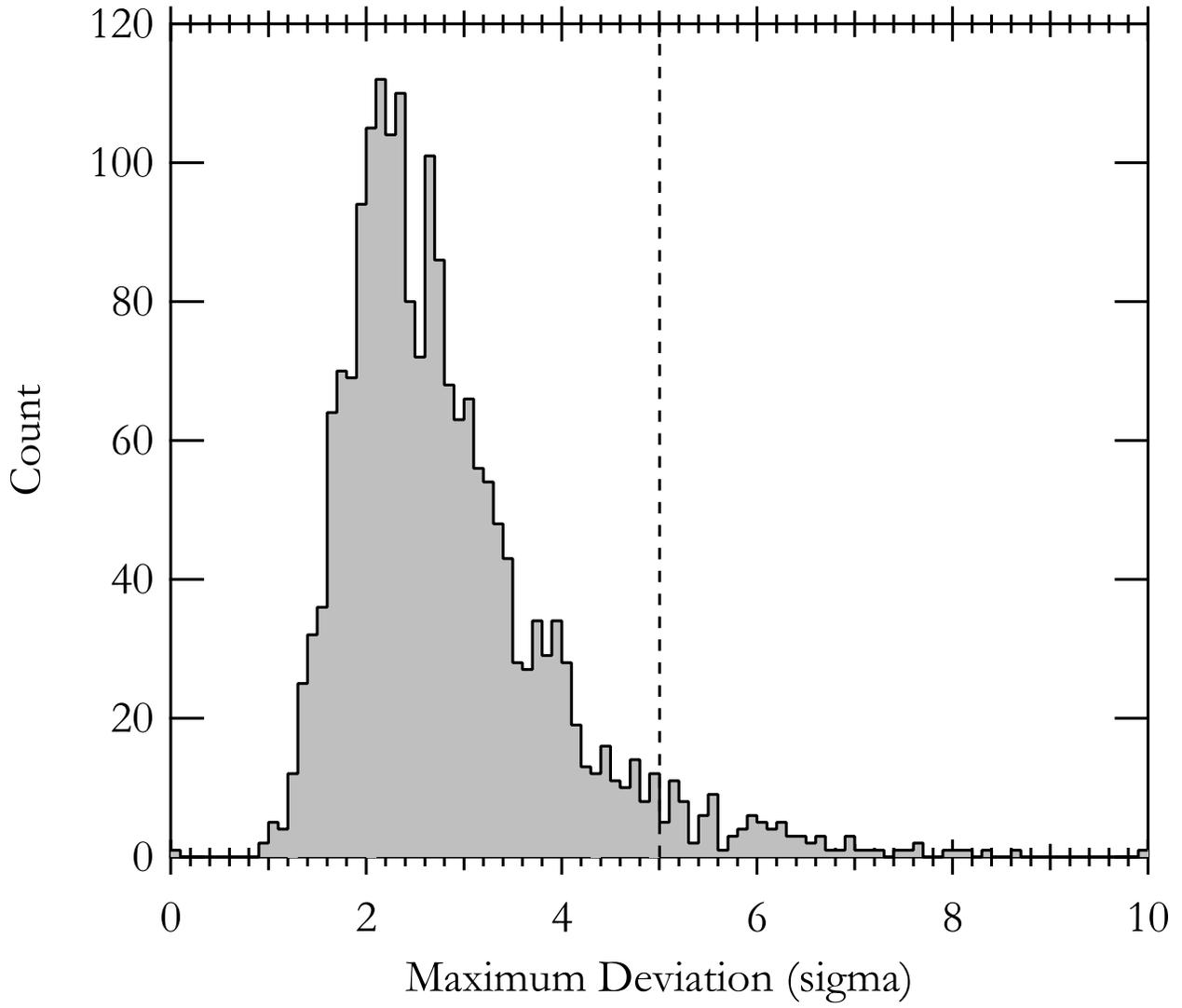}
  \end{center}
  \caption{The distribution of maximum deviation of number density measured in
the unit of a standard deviation at radial distance from 1~Mpc to $r_{\mbox{\scriptsize bg,min}}$
for 1976 AGN samples. The vertical dashed line represent the maximum value for 
sample selection.}
  \label{fig:devmax}
\end{figure*}

It is also possible that the nearby cluster or high density region is located
in a region offset from the AGN.
To remove such samples, the reduced $\chi^{2}$ relative to the flat distribution
and maximum deviation $\sigma_{\mbox{\scriptsize max}}$ of radial distribution of galaxy density 
from the average was calculated at distance range of 1~Mpc to $r_{\mbox{\scriptsize bg,min}}$ .
The average density was calculated at distance range from $r_{\mbox{\scriptsize bg,min}}$ to 
$r_{\mbox{\scriptsize bg,max}}$.
The error used for the calculation of the $\chi^{2}$ is statistical one.
The distributions of the reduced $\chi^{2}$ and maximum deviation  $\sigma_{\mbox{\scriptsize max}}$ 
for the 1,976 AGN samples are shown in figures~\ref{fig:chi2} and \ref{fig:devmax}.
The criteria used here were $\chi^{2}/(n-1) \le 3.0$ and $\sigma_{\mbox{\scriptsize max}} \le 5.0$.
After this selection, the total number of AGNs was 1,828.

Figure~\ref{fig:M_z} shows the distribution of K-corrected V-band absolute magnitude 
and redshift of the AGNs selected by the above criteria.
Open circles represent AGNs for which the analysis was made using the UKIDSS data,
and crosses represent AGNs for which the analysis was made using the Suprime-Cam data.
The histograms of AGN redshift for dim AGNs ($M_{V} \ge -25$) and bright 
AGNs ($M_{V} < -25$) are shown in figure~\ref{fig:zhist_D}.
The low redshift ($z<0.6$) and high luminosity ($M_{V} < -25$) AGNs, 
and high redshift ($z\ge1.3$) and low luminosity ($M_{V} \ge -25$)
AGNs were not analyzed for their poor statistics.
As a result the total number of analyzed AGNs was reduced to 1,809.

In table~\ref{tbl:number_AGN}, the number of AGNs used in this work are
shown for each data subset (z1-D to z5-B defined in section~\ref{sec:analysis}) 
and for each origin of the AGN identification.
The column ``SDSS'' of the table represents the number of AGNs which are contained in the
SDSS QSO catalog (4-th ed.).
The column ``2dF'' represents the number of AGNs which are contained in the 2dF QSO catalog 
but not contained in the SDSS catalog.
The AGNs which are contained in both the SDSS and 2dF catalog are counted as ``SDSS''.
Among the AGNs which are not contained in the SDSS catalog nor the 2dF catalog, the 
AGNs which were identified based on an optical, UV, or infrared observations are counted as 
``UV-OPT-IR'', the AGNs which were selected from X-ray or Radio catalog are
counted as ``XRAY'' or ``RADIO'', respectively.
About 90\% of the AGNs are selected from the large survey of SDSS and 2dF, thus
our AGN samples are nearly homogeneous and dominated by optically selected AGNs.

The celestial distribution of the 1,809 AGNs is shown in figure~\ref{fig:skymap}.
Since a large fraction of AGNs comes from the SDSS catalog, the distribution 
almost follows the SDSS survey area.
The distribution of the right ascension and the declination are also shown in 
figure~\ref{fig:coord_hist} for all the samples (open histogram)
and samples which have data of Suprime-Cam (shading histogram).
The concentration to declination $\sim 0$ degree is due to the 
contribution of UKIDSS LAS survey.
The most significant concentration of Suprime-Cam observations is located at
the Sextans field (RA$=150^{\circ}$ -- $157^{\circ}$, Dec$=-4^{\circ}$ -- $+1^{\circ}$).
The total number of AGNs in the Sextans field is 80,
and their number fraction at each subset group is $\sim$ 20\% for the z4-B
and z5-B group, and less than 6\% for the other groups.
We confirmed that this concentration to a specific field did not affect the result
so much, and it will be discussed in section~\ref{sec:result}.
%%
%% Thus the bias to this specific field is 
%%
Although there is a slight concentration to specific fields,
the AGNs are distributed in an extended area of the sky, 
and we can expect that the bias effect (cosmic variance) would be mostly 
negligible in our analysis.
%%

%%
%% データの統計
%%
%% In Table~\ref{tbl:stat_band},
%% the numbers of AGN samples, $n_{AGN}$, and
%% the minimum, maximum and median value of the limiting magnitude
%% are summarized 
%% for samples observed in both of the optical and infrared bands,
%% only in the optical bands,
%% and only in the infrared band.
%%
%% 23\% of the sample were observed in both the optical and infrared bands,
%% 59\% were observed in the optical band only, and
%% 18\% were observed in the infrared band only.
%%
%% The typical limiting magnitudes for optical and infrared bands were
%% 24.8 and 19.6, respectively.
%%

%%
%% Suprime-Cam 画像データについて
%%
The Suprime-Cam images provided by the JVO portal were processed using the JVO 
Suprime-Cam reduction pipeline (version 0.1.24) which was developed based 
on the Suprime-cam Deep Field REDuction (SDFRED) package (\cite{Ouchi_2004}).
The data reduction was carried out on the JVO grid computing system (\cite{Shirasaki_2008}).
All the data obtained before December 2006 were used to create mosaic images.
The SDFRED is based on a software package developed for Suprime-Cam data reduction
(\cite{Yagi_2002}).
The data were reduced in the following procedure for each CCD frame:
1) bias subtraction;
2) flat field correction;
3) distortion correction; and
4) astrometric correction.
The positions, rotation angles, and flux normalization factors
relative to a reference frame were calculated
for every frame, then they were stacked to make a single mosaic image.
%%
%% フォトメトリーと位置の精度
%%
The photometry was performed using the data of standard star observations 
which were taken on the day that one of the stacked frames were observed.
The comparison with the SDSS catalog shows the differences of photometric
magnitudes were less than 0.2 for 90\% of the data compared.
The astrometric accuracy was estimated by comparing with the USNO catalog.
The peak of the distribution of angular difference from the USNO data was
0.3 arcsec, and the difference was less than 1.2 arcsecond
for 99\% of the analysed data.
%%

% \clearpage

%
% rho_0 の評価
%
\section{Estimation of $\rho_{0}$}
\label{sec:rho0}

Estimation of $\rho_{0}$, which is an average number density of observed
galaxy at AGN redshift, is crucial for determining the cross correlation 
length of AGNs  and galaxies.
Since redshift of galaxy is not measured in this work, $\rho_{0}$ cannot be 
determined directly from the data used here.
Therefore, it needs to be indirectly derived by model calculation under
reasonable assumptions.
The luminosity function of galaxy $\phi(M)$ is one of the functions
required for the model calculation.
The only observable that can be used to infer the density of observed galaxy
is an observed magnitude distribution $N(m)$.
The observed magnitude distribution is a product of true magnitude distribution
$N_{\mbox{\scriptsize true}}(m)$ and detection efficiency $DE(m)$ for an object of apparent 
brightness $m$.
Once the model functions for $\phi(M)$, $N_{\mbox{\scriptsize true}}(m)$ and $DE(m)$ 
are derived, we can estimate $\rho_{0}$ by comparing the observable $N(m)$ with 
the model prediction.
For the UKIDSS LAS data, however, model parameters of $DE(m)$ are not well 
determined from observation because of poor statistics.
Thus the parameter, $m_{\mbox{\scriptsize limit}}$, which characterizes $DE(m)$ 
is used to estimate $\rho_{0}$.

An outline of the procedure for determining $\rho_{0}$ and 
its uncertainty is described as follows;
we define a parameter $m_{\mbox{\scriptsize limit}}$ which is closely related with
the limiting apparent magnitude of observation and is used to calculate
$\rho_{0}$ by integrating the luminosity function $\phi(M)$ to the 
corresponding absolute magnitude at the AGN redshift.
We estimate $m_{\mbox{\scriptsize limit}}$ under the assumption than the cumulative 
fraction of apparent magnitude above $m_{\mbox{\scriptsize peak}}$ has a constant 
value of $F_{\mbox{\scriptsize limit}}$ at $m_{\mbox{\scriptsize limit}}$, where $m_{\mbox{\scriptsize peak}}$
is a peak magnitude of the observed apparent magnitude distribution.
This assumption would be valid if the observations were carried out under 
identical conditions.
This is, however, not actually the case, and $F_{\mbox{\scriptsize limit}}$ is
not a constant but a variable depending on the observational condition.
Thus we calculated the expected value of $F_{\mbox{\scriptsize limit}}$ for 
each plausible observational condition, and we considered their average to 
be an estimator for the most probable value of $\rho_{0}$, and the lower and 
upper limit of $F_{\mbox{\scriptsize limit}}$ to be an estimator for uncertainty 
of $\rho_{0}$.
The detail of the procedure is described below.
$\rho_{0}$ is calculated by integrating the product of the luminosity function 
$\phi(M; z, \lambda)$ and detection efficiency $DE(M+DM)$ for a source with
absolute brightness of $M$ (apparent brightness $m = M + DM$) at AGN redshift $z$.
$DM$ is a distance modulus for redshift $z$.
We use an effective limiting magnitude $m_{\mbox{\scriptsize limit}}$ defined below
rather than directly using $DE(m)$.
This is because $DE(m)$ is not necessarily determined accurately from the data 
especially for data with poor statistic, thus we decided to define a parameter 
which is robust even for a sample with small number of observed objects.
$m_{\mbox{\scriptsize limit}}$ is defined so that the integral of luminosity function up 
to $m_{\mbox{\scriptsize limit}} - DM$ equals to $\rho_{0}$:
\begin{eqnarray}
   \rho_{0} &=& \int_{m_{\mbox{\scriptsize low}}-DM}^{\infty} \phi (M) DE (M+DM) dM \nonumber \\
            &=& \int_{m_{\mbox{\scriptsize low}}-DM}^{m_{\mbox{\scriptsize limit}}-DM} \phi (M) dM,
\end{eqnarray}
where $m_{\mbox{\scriptsize low}}$ is a lower boundary of the apparent magnitude 
for the integration.

\begin{figure*}
  \begin{center}
      \FigureFile(\textwidth,\textwidth){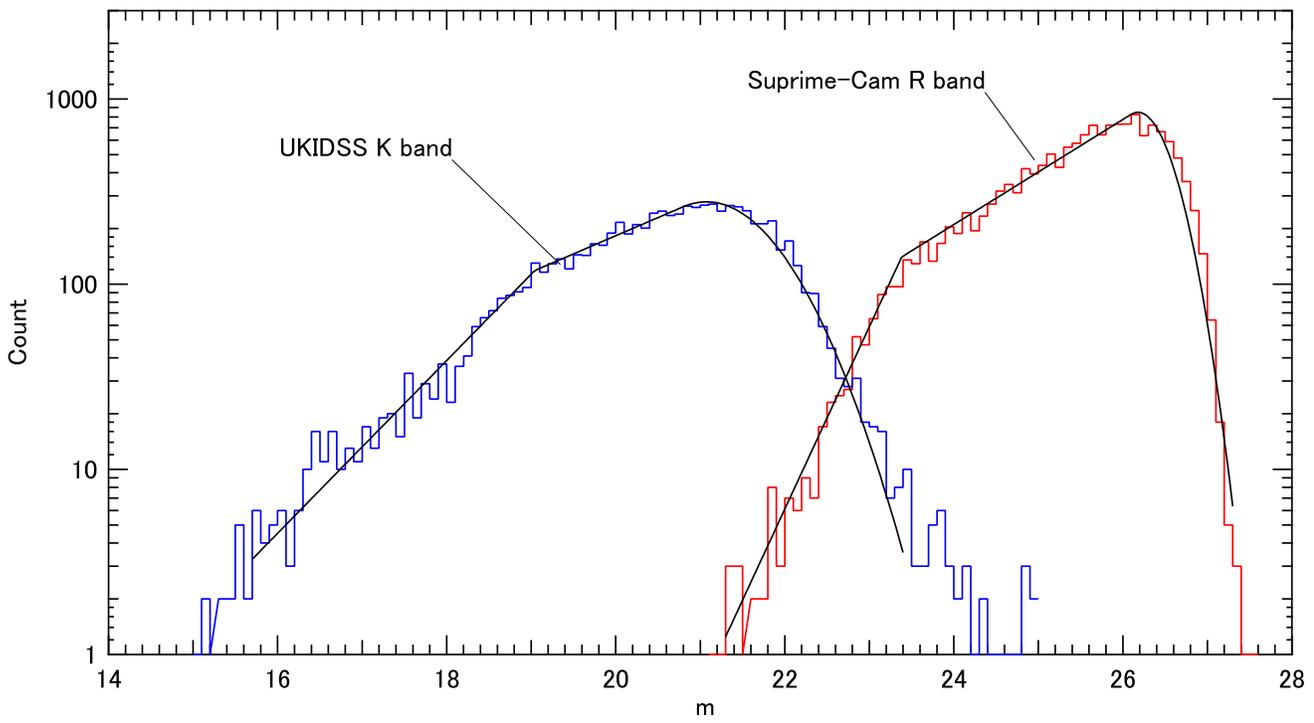}
  \end{center}
  \caption{Examples of the observed magnitude distributions and their model functions.
The left histogram shows the UKIDSS K-band data around the AGN ``XMM~J02181-0438''
and the right one shows the Suprime-Cam R-band data around the AGN ``US~2694''.}
  \label{fig:MagDist_RK}
\end{figure*}

To derive a relation for determining $m_{\mbox{\scriptsize limit}}$ from  observational
data, we performed simple numerical calculation as follows:
%%
%% First we describe the model of luminosity function used in this work.
%%
We parametrized the luminosity function of galaxy by means of the absolute magnitude
$M$, redshift $z$, and source frame wavelength $\lambda$ by using the luminosity
functions obtained by
\citet{Gabasch_2004} for 1500\AA, 2800\AA, u, B and g' bands,
\citet{Gabasch_2006} for r', i', and z' bands, 
and \citet{Cirasuolo_2007} for K band.
The parameters $\alpha$, $\phi_{0}$, and $M_{0}$ of the Schechter function for each rest-frame
wavelength band were parametrized by a polynomial function of redshift.
Then, for each redshift, these parameters were represented as a function of
rest-frame wavelength by interpolating with the cubic spline.
Using this parametrization, we obtained luminosity function $\phi(M; z, \lambda)$ 
for an arbitrary redshift and rest-frame wavelength.
The luminosity functions were evaluated for the AGN redshift and for the rest frame 
wavelength corresponding to the center wavelength of the observation band.
The uncertainty of the luminosity function itself is 5\% to 10\% for redshift less 
than 1.6, and 20~\% around $z$ $\sim$ 2.0.
As explained below, we overestimate the uncertainty of $m_{\mbox{\scriptsize limit}}$
to derive a conservative estimate, we can expect that the error caused by the 
uncertainty of the luminosity function is inclusive in our error estimate of $\rho_{0}$.
Even if this is not the case, the additional error to $r_{0}$ is less than 10\%
of the error estimated from Poisson statistic and uncertainty of 
$m_{\mbox{\scriptsize limit}}$.
We assumed the following function for modeling the detection efficiency :
\begin{eqnarray}
   DE(m)  = \left\{
   \begin{array}{ll}
     1  & ( m < m_{\mbox{\scriptsize th}} )  \\
     \exp{( -(m-m_{\mbox{\scriptsize th}})^2/\sigma_{m}^{2} ) }  & ( m \ge m_{\mbox{\scriptsize th}}), \\
   \end{array}
   \right.
   \label{eq:det_eff}
\end{eqnarray}
where $m_{\mbox{\scriptsize th}}$ and $\sigma_{m}$ 
represent the magnitude where detection efficiency starts to decrease and 
the attenuation width of the detection efficiency, respectively.
In the case of Suprime-Cam data, the detection efficiency rapidly decreases at the bright
magnitude side owing to the masking procedure described in the previous section.
This effect was approximated by introducing the lower cut-off magnitude $m_{\mbox{\scriptsize low}}$.
%%
%% $m_{\mbox{\scriptsize low}}$ was determined so that 5\% of the total number of observed sources 
%% are below this magnitude.
%%
$m_{\mbox{\scriptsize low}}$ is a minimum of the observed magnitude excluding the 
first 5\% of all the sources ordered with their magnitudes.
The number density at the bright end of the luminosity function is small and the 
error on the $\rho_{0}$ estimation caused by uncertainty of $m_{\mbox{\scriptsize low}}$
is expected to be lower than a few \%, which is
negligible compared to the error caused from the uncertainty of $m_{\mbox{\scriptsize limit}}$
which we estimate to be 20 to 40\%.
The only observable that can be used to determine $m_{\mbox{\scriptsize limit}}$ is 
the observed magnitude distribution $N(m)$.
Especially, the distribution beyond the peak magnitude $m_{\mbox{\scriptsize peak}}$ 
can be a sensitive estimator for $m_{\mbox{\scriptsize limit}}$.
Thus we define the cumulative fraction $F(m)$ to estimate $m_{\mbox{\scriptsize limit}}$
as follows:
\begin{equation}
F(m) = \int_{m_{\mbox{\scriptsize peak}}}^{m} N(m') dm' / 
       \int_{m_{\mbox{\scriptsize peak}}}^{\infty} N(m') dm'.
\label{eq:cumulative_fraction}
\end{equation}
%%
%% where $m_{\mbox{\scriptsize max}}$ is a maximum magnitude of the detected sources.
%%
To calculate a model function for $F(m)$, we need a model function for {\it true} 
magnitude distribution $N_{\mbox{\scriptsize true}}(m)$.
We assumed a broken power law function for $N_{\mbox{\scriptsize true}}(m)$:
\begin{eqnarray}
   N_{\mbox{\scriptsize true}}(m)  = \left\{
   \begin{array}{ll}
     c \cdot 10^{a (m-m_{\mbox{\scriptsize b}})}    & ( m < m_{\mbox{\scriptsize b}} )  \\
     c \cdot 10^{b (m-m_{\mbox{\scriptsize b}})}    & ( m \ge m_{\mbox{\scriptsize b}}), \\
   \end{array}
   \right.
   \label{eq:mag_dist_org}
\end{eqnarray}
where $m_{\mbox{\scriptsize b}}$ is a break magnitude of the magnitude distribution.
Using this model function, the observed magnitude distribution can be calculated
as $N_{\mbox{\scriptsize model}}(m) = N_{\mbox{\scriptsize true}}(m) \times DE(m)$.
Applying this to the equation~(\ref{eq:cumulative_fraction}),
we can obtain the model function $F_{\mbox{\scriptsize model}}(m)$.
To show how well the model function $N_{\mbox{\scriptsize model}}(m)$ describes an
actual observed magnitude distribution, the model functions fitted to the observation
are shown in figure~\ref{fig:MagDist_RK}.
The fitting results for two examples of UKIDSS K-band (left histogram) and 
Suprime-Cam R-band (right) are shown in the figure.
The best fit parameters for these examples are:
$a=0.47\pm0.02$, $b=0.20\pm0.01$, $m_{\mbox{\scriptsize b}}=19.0\pm0.1$, 
$m_{\mbox{\scriptsize th}} = 20.8\pm0.06$ and
$\sigma_{m} = 1.11\pm 0.03$ for UKIDSS K-band, and
$a=0.98\pm0.05$, $b=0.28\pm0.01$, $m_{\mbox{\scriptsize b}}=23.4\pm0.04$, 
$m_{\mbox{\scriptsize th}} = 26.1\pm0.02$ and
$\sigma_{m} = 0.51\pm0.01$ for Suprime-Cam R-band.
The break magnitude $m_{\mbox{\scriptsize b}}$ approximately corresponds to $M_{0}$ of the Schechter
function for a maximum observable redshift, thus it is expected to correlate with 
$m_{\mbox{\scriptsize th}}$ which is a parameter related with sensitivity of the observation.
Since an accurate value of $m_{\mbox{\scriptsize b}}$ is not required for the estimation of 
$m_{\mbox{\scriptsize limit}}$ and since most AGN field have poorer statistics, when
fitting $N(m)$ for all the AGN samples
we simply used an empirical relation of $m_{\mbox{\scriptsize b}} = m_{\mbox{\scriptsize th}} - 2.5$,
which is deduced from the fitting result shown above.
The cumulative fraction at $m_{\mbox{\scriptsize limit}}$, $F_{\mbox{\scriptsize limit}} = 
F(m_{\mbox{\scriptsize limit}})$,
depends on the limiting magnitude of the observation, redshift of the AGN, the 
observation band, and the model parameter of $N_{\mbox{\scriptsize true}}$.
We, therefore, calculated $F_{\mbox{\scriptsize limit}}$ for various plausible parameter
ranges:
$z =$ 0.3--3.0,
$\sigma_{m} =$ 0.5--1.2,
$a =$ 0.6,
$b =$ 0.1 -- 0.6, 
$m_{\mbox{\scriptsize th}} =$ 19 -- 26, 
and $\rho_{0} > 10^{-5}$.
The last criteria is the same as was applied to the observational data.
%%
%% In Figure~\ref{fig:f_vs_mpeak}, $F_{\mbox{\scriptsize limit}}$ calculated for
%% several parameter sets are shown as a function of $m_{\mbox{\scriptsize peak}}$.
%%
%% The value of $F_{\mbox{\scriptsize limit}}$ is sensitive to parameters $b$ and $\sigma_{m}$.
%%
%% For a typical paramter of $b =$ 0.2--0.3 and $\sigma_{m} =$ 0.5--1.0, 
%% $F_{\mbox{\scriptsize limit}}$ is in the range of 0.6--0.8.
%%
%%
To choose the ranges of parameters $\sigma_{m}$, $b$ and $m_{\mbox{\scriptsize th}}$, we fit the
magnitude distribution for each AGN sample to the model function $N_{\mbox{\scriptsize model}}(m)$.
The statistics of UKIDSS LAS samples were too poor to fit to the model function, thus
we performed the fitting only for datasets of $m_{\mbox{\scriptsize peak}} > 21$.
%%
%% we 
%% summed every five samples in order from the lowest $m_{\mbox{\scriptsize peak}}$ sample, 
%% and fit the model to the summed distribution.
%%
The ranges of $b$, $\sigma_{m}$ and $m_{th}$ were determined so that more than 90\%
of the samples were included.
Since $F_{\mbox{\scriptsize limit}}$ is insensitive to the parameter $a$, $a$ was fixed to
an average value of 0.6.
For these parameter ranges, we obtained 0.1, 0.6, and 0.8 for the minimum, median, and
maximum values of $F_{\mbox{\scriptsize limit}}$, respectively.
Following this result, we calculated the most probable value of $\rho_{0}$ by integrating
the luminosity function up to $m_{\mbox{\scriptsize limit}}$ determined with 
$F_{\mbox{\scriptsize limit}} =$ 0.6, and lower and upper boundary with
$F_{\mbox{\scriptsize limit}} =$ 0.1 and 0.8, respectively.
The error estimated in this way could be an overestimation, but in this work
we used the error as a conservative estimate.
%%

% \clearpage

%%
%%
%%
\section{Analysis}
\label{sec:analysis}

%\subsection{Cross-Correlation Method}

%%
%% 2点相関係数
%%
The two-point cross-correlation function, $\xi$, is generally used to measure the 
clustering between an AGN and its surrounding galaxies. It is defined as:
\begin{equation}
   \xi(r) = \rho(r)/\rho_{0} - 1,
   \label{eq:xi}
\end{equation}
where $\rho(r)$ is the number density of galaxies at a distance $r$ from an AGN,
and $\rho_{0}$ is the average number density of galaxies at the AGN redshift.
Since the redshifts were not measured for galaxies, the projected 
cross-correlation function, $\omega$, was derived from the observational data in the
following way:
The projected cross-correlation function is obtained by integrating the
equation~(\ref{eq:xi}) along the line of sight, then it is described as
(\cite{Davis_1983}):
\begin{equation}
   \omega(r_{p})  =  2 \int_{0}^{\infty} \xi(r_{p}, \pi) d\pi
                 =  2 \int_{r_{p}}^{\infty} r dr \xi (r) (r^{2} - r_{p}^{2})^{-0.5},
\label{eq:omega1}
\end{equation}
where $r_{p}$ and $\pi$ are components of distance from the AGN 
perpendicular to the line of sight and along the line of sight,
respectively.
We assumed a power law function for the density profile around an AGN,
i.e. $\xi(r) = (r_{0} / r)^{\gamma}$.
In this case, equation~(\ref{eq:omega1}) can be integrated analytically as:
\begin{equation}
   \omega(r_{p}) = r_{p} \left( \frac{r_{0}}{r_{p}} \right)^{\gamma}
               \frac{\Gamma (1/2) \Gamma ( (\gamma - 1) / 2 ) }{ \Gamma(\gamma/2)},
   \label{eq:omega_2}
\end{equation}
where $\Gamma$ is the Gamma function.

Studies of cross-correlation analysis usually use a method which counts pairs of 
objects with a given separation and compares with the expected background which 
is a number of pairs in a random sample for the same separation, 
and these counts are denoted as $DD(r)$ and $DR(r)$ respectively.
The cross-correlation function is calculated as (\cite{Davis_1983}):
\begin{equation}
   \xi(r) = \frac{DD(r)}{DR(r)} - 1.
   \label{eq:corr_estimator}
\end{equation}
In this work, however, data of each AGN field are highly inhomogeneous, thus it is
not appropriate to use the random catalog created from the whole dataset
for estimating the background level.
In addition, we need to consider two kinds of background.
One is a background corresponding to $DR(r)$ in the equation~(\ref{eq:corr_estimator}),
which is an average number density at the AGN redshift.
We estimate this background $\rho_{0}$ as explained in section~\ref{sec:rho0}.
The other background is an integrated number density of all the observed galaxies
from $z = 0$ to infinity, which needs to be considered in our case as the redshifts 
of galaxies were not measured in our sample.
The second background, $n_{\mbox{\scriptsize bg}}$, is estimated by fitting 
model function derived below (equation~\ref{eq:model}) to the observed data.
In addition, the coordinates of AGNs in our samples are well separated from each
other, and the cross-correlation function for each AGN can be calculated independently
from the other AGN samples.
Thus, in our case, it is enough to simply calculate the number density around each 
AGNs and to take their average.

The projected correlation function $\omega(r_{p})$ is calculated using the projected
number density $n(r_{p})$ as:
\begin{eqnarray}
   \omega(r_{p}) & = & \int_{-\infty}^{\infty} \xi(r_{p}, \pi) d\pi
                = \frac{1}{\rho_{0}} \int_{-\infty}^{\infty} (\rho(r) - \rho_{0}) d\pi \nonumber \\
                & = & \frac{n(r_{p}) - n_{\mbox{\scriptsize bg}}}{\rho_{0}},
\end{eqnarray}
where the integral of $\rho(r)$ and $\rho_{0}$ are substituted with $n(r_{p})$ and
the surface density of background galaxies $n_{\mbox{\scriptsize bg}}$, respectively;
\begin{equation}
  \int_{-\infty}^{\infty} \rho(r) d\pi = n(r_{p}),
\end{equation}
\begin{equation}
  \int_{-\infty}^{\infty} \rho_{0} d\pi = n_{\mbox{\scriptsize bg}}.
\end{equation}
An accurate estimate of $n_{\mbox{\scriptsize bg}}$ is crucial for deriving the
correlation function correctly, however, it is not possible to determine
$n_{\mbox{\scriptsize bg}}$ with good enough precision for each AGN sample.
Thus the average of the background $<n_{\mbox{\scriptsize bg}}>$ was estimated
from the average of the projected number density $<n(r_{p})>$, and
the projected correlation function averaged for $m$ AGN were calculated
using the following formula rather than calculating the average of $\omega(r_{p})$ 
for each AGN:
%%
%% by dividing the excess number density calculated by subtracting
%% background density from the average of projected number density $<n(r_{p})>$
%% by average of $\rho_{0}$
%% rather than calculating the average of $\omega(r_{p})$ for each AGN as follows:
%%
\begin{equation}
   \omega(r_{p}) = \frac{<n(r_{p})> - <n_{\mbox{\scriptsize bg}}>}{<\rho_{0}>},
   \label{eq:omega_3}
\end{equation}
where $<\rho_{0}>$ is the average of $\rho_{0}$ determined for each AGN sample.

%%
%% We measured number density of all the detected sources around
%% AGNs of a given redshift range.
%%
The number density at perpendicular distance $r_{p}$ from the i-th AGN is 
calculated as:
\begin{equation}
   n_{i}(r_{p}) = N_{i}(r_{p})/S_{i}(r_{p}),
\end{equation}
where $N_{i}(r_{p})$ is the number of detected sources at a distance range
of $r_{p}$ to $r_{p} + dr$ for the $i$-th AGN, and $S_{i}(r_{p})$ is the projected
area corresponding to the distance range.
In calculating $N_{i}(r_{p})$, we took into account the coverage
fraction for each annulus $f_{i}(r_{p})$, which is discussed in
section~\ref{sec:dataset}, by multiplying $1/f_{i}(r_{p})$ to
a number of objects detected at a distance bin $r_{p}$ to $r_{p} + dr$.
Then the average of $n(r_{p})$ and $\rho_{0}$ over $m$ AGNs are given by:
\begin{equation}
   <n(r_{p})> = \frac{1}{m} \sum n_{i}(r_{p}),
\end{equation}
%%%%%%%
\begin{equation}
   <\rho_{0}> = \frac{1}{m} \sum \rho_{i,0},
\end{equation}
%%
%% where $n_{bg}$ is the number density integrated along the line 
%% of sight at distance large enough for the clustering due to the
%% AGN to becomes negligible.
%%
where $\rho_{i,0}$ is the averaged number density of observed galaxies 
at the i-th AGN redshift.
$\rho_{i,0}$ was calculated for the rest frame wavelength corresponding
to each observation band as described in section~\ref{sec:rho0}.
The rest frame wavelength corresponding to the observation band
was calculated by $\lambda_{\mbox{\scriptsize rest}} = \lambda_{\mbox{\scriptsize center}} / (1 + z)$,
where $\lambda_{\mbox{\scriptsize center}}$ is the effective center wavelength
of the observation band.

\begin{figure*}
   \begin{center}
       \FigureFile(\textwidth,\textwidth){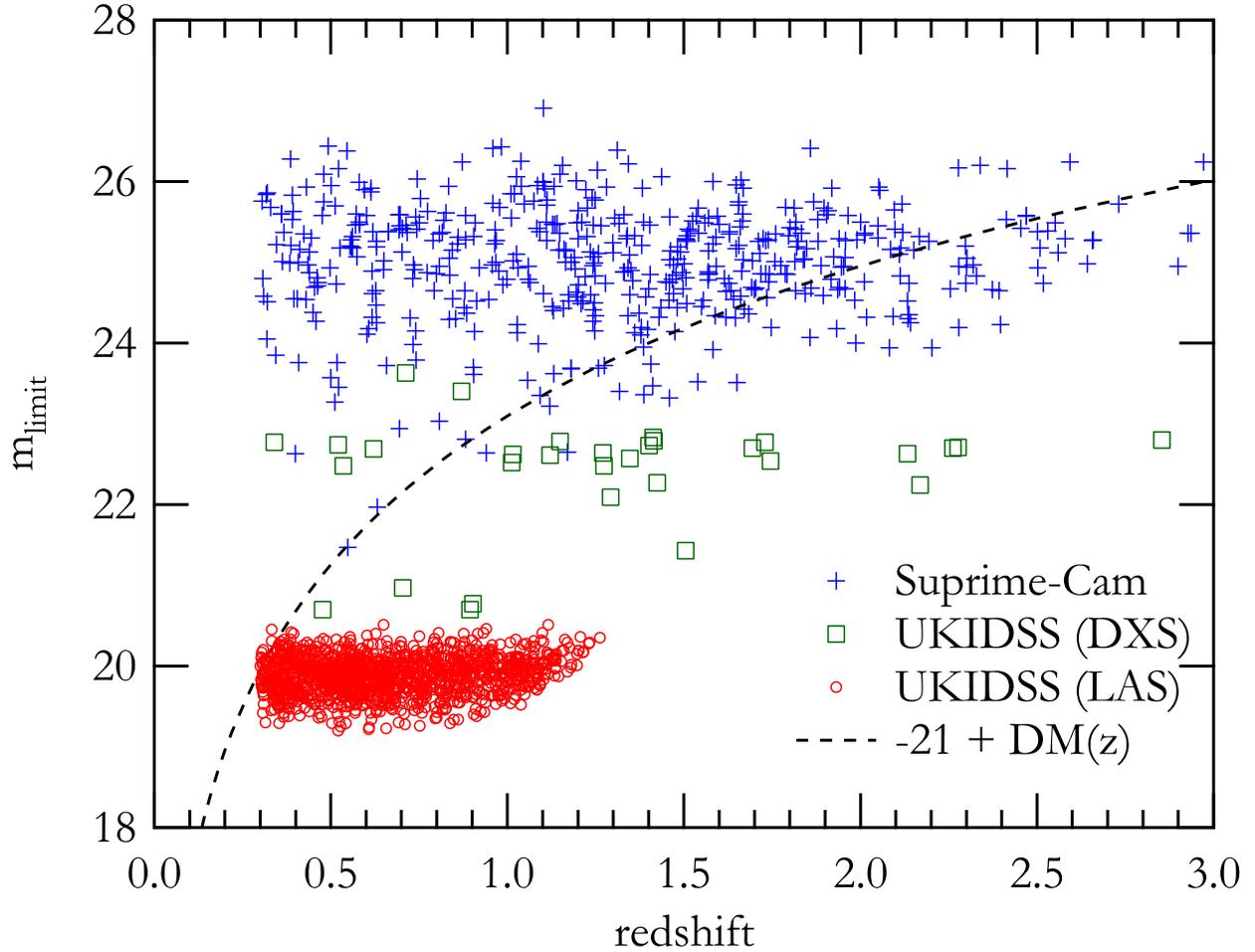}
  \end{center}
  \caption{Distribution of $m_{\mbox{\scriptsize limit}}$ and AGN redshift.

The open circles and boxes correspond to AGNs for which the galaxy sample is derived 
from the data of UKIDSS Large Area Survey (LAS) and Deep Extragalactic Survey (DXS), 
respectively.
The crosses correspond to AGNs for which the galaxy sample is derived from the
Suprime-Cam data. The dashed line represents apparent magnitude which corresponds to
absolute magnitude of $-21$ at the redshift.}
  \label{fig:mlimit_vs_z}
\end{figure*}

We used only the data observed at the most sensitive band for detecting the
galaxies at the AGN redshift.
The observation band with the largest value of $\rho_{i,0}$ was chosen as
the most sensitive band.
To show the depth of observations used in this work, $m_{\mbox{\scriptsize limit}}$ of the 
observation with the most sensitive band is plotted against redshift in
figure~\ref{fig:mlimit_vs_z}.
Open circles (squares) represent AGNs for which UKIDSS LAS (DXS) survey data were 
used, and crosses represent AGNs for which Suprime-Cam data were used.
The apparent magnitude corresponding to an absolute magnitude of $-21$ is shown as
a dashed line.
The number of AGN samples for each observation band are summarized in
table~\ref{tbl:number_band}.
%%
%%

%%
%% The galaxy number count, $N_{i}(r_{p})$, was derived from the multi-band catalog.
%%
Galaxies near an AGN may be contaminated by the light of the AGN, and the detection
efficiency for galaxies might be lowered there.
Thus, in calculating $N_{i}(r_{p})$, the region of 4 arcsec from the AGN was masked.
The typical FWHM of the Suprime-Cam image used in this work is less than 1~arcsec,
therefore the contamination effect of the AGN light to the galaxy detection efficiency 
would be negligible outside the masked region.
%%
%% if the apparent brightness of the AGN is dimmer than 17~mag.
%%
The reduction of the projected area due to the masking was taken into
account in the calculation of $S_{i}(r_{p})$.

%%%
%%%
%%%
%%% \subsection{Sample Division}

%%
%% データセットの分類 
%%
We divided all the AGN samples into five redshift groups, 
$z =$ 0.3--0.6 (referred to as z1), 0.6--0.9 (z2), 0.9--1.3 (z3), 
1.3--1.8 (z4) and 1.8--3.0 (z5).
For each redshift group, the dataset is further divided into a dimmer group
($M_{V} \ge -25$, referred by putting a suffix D after the redshift group name, 
e.g. z1-D) and a brighter group ($M_{V} < -25$, suffix B).
In figure~\ref{fig:M_z}, the redshift and luminosity range of each group 
is shown with a solid box together with the absolute magnitude and redshift 
distribution of the AGNs.
Hereafter we call an AGN of $M_{V} \ge -25$ as ``dim AGN'', and an AGN of 
$M_{V} < -25$ ``bright AGN''.
The bright AGNs for $z < 0.6$ and the dim AGNs for $z \ge 1.3$ were
not analyzed due to their poor statistics.
For each group the averaged galaxy number density, $<n(r_{p})>$, was calculated as a 
function of perpendicular comoving distance from the AGN, and it was fitted with 
a model function.
The model function is derived from equations~(\ref{eq:omega_2}) and (\ref{eq:omega_3})
as follows:
\begin{equation}
   <n(r_{p})> = r_{p} \left( \frac{r_{0}}{r_{p}} \right)^{\gamma}
               \frac{\Gamma (1/2) \Gamma ( (\gamma - 1) / 2 ) }{ \Gamma(\gamma/2)} 
               <\rho_{0}>
              + <n_{bg}>.
   \label{eq:model}
\end{equation}
%%

\begin{figure*}
  \begin{center}
      \FigureFile(0.85\textwidth,\textwidth){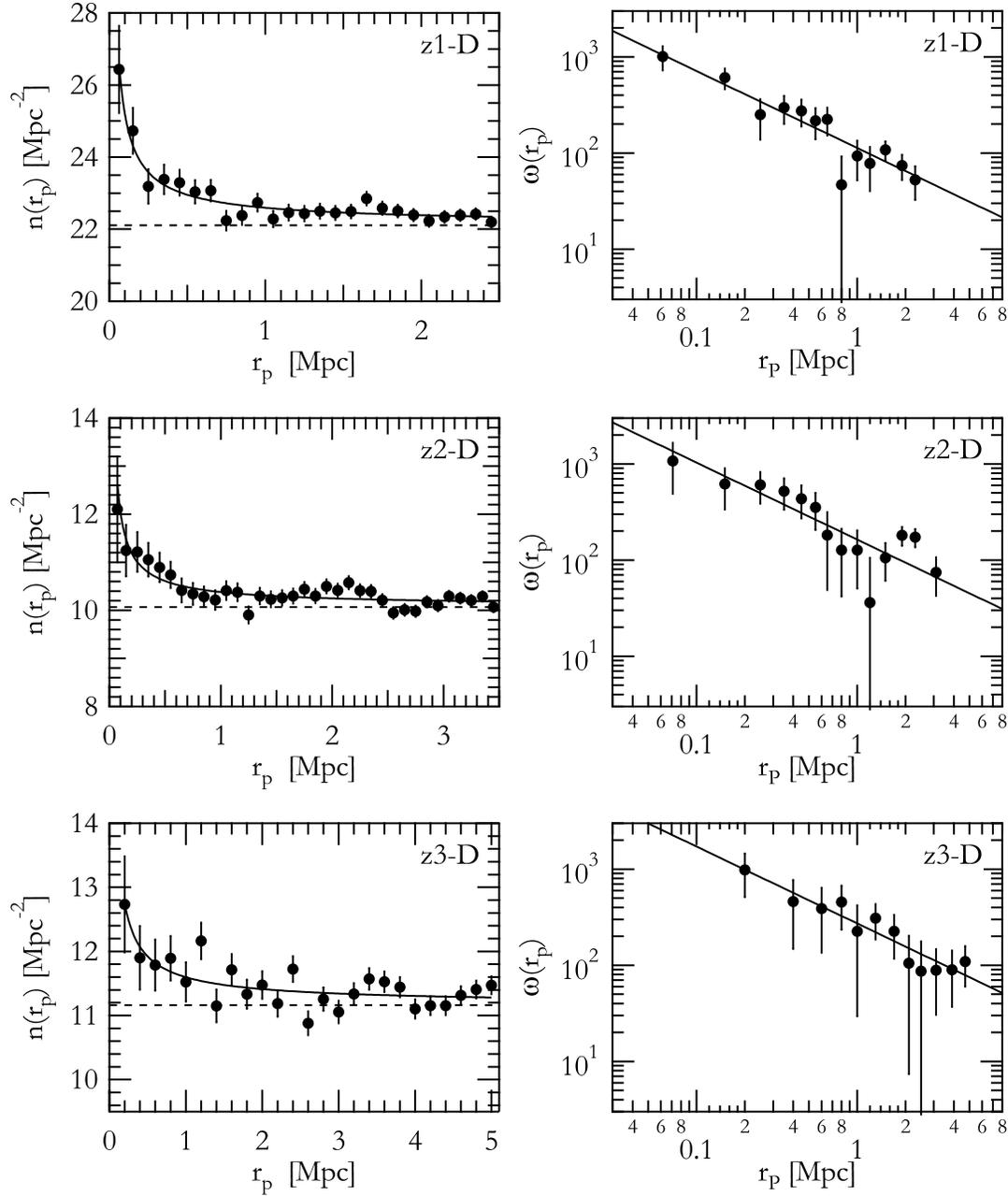}
  \end{center}
  \caption{The distributions of number density of detected sources and the projected 
correlation functions for dim AGN groups. 
The left panels shows the averaged galaxy number density 
as a function of projected distance from the AGN, and the corresponding projected
correlation functions are shown in the right panels.
The data points of the right panels were calculated by rebinning the data points of
left panels to be shown with roughly even bins in $\log{r_{p}}$.
The dashed lines in the left panels represent fitting parameter $n_{\mbox{\scriptsize bg}}$.
The solid lines represent the model function fitted to the observational data.
}
  \label{fig:density_dim}
\end{figure*}

\begin{figure*}
  \begin{center}
      \FigureFile(0.85\textwidth,\textwidth){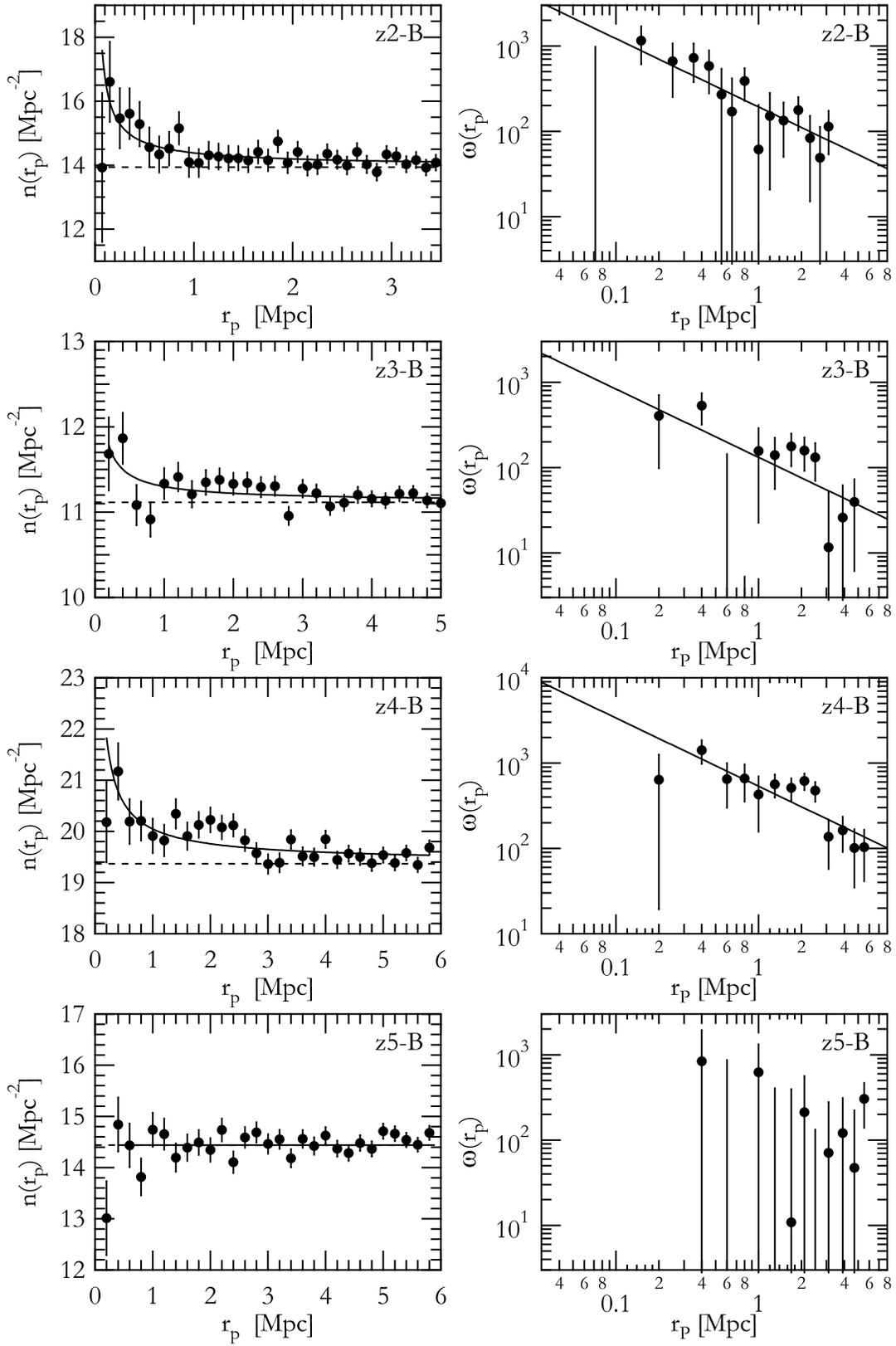}
  \end{center}
  \caption{The distributions of number density of detected sources and the projected 
correlation functions for bright AGN groups. The other explanation of this figure
is the same as figure~\ref{fig:density_dim}.
}
  \label{fig:density_bright}
\end{figure*}

%%
%% 数密度分布の図を説明
%%
In the left panels of figures~\ref{fig:density_dim} and \ref{fig:density_bright} the 
galaxy number densities for each AGN group are shown with closed circles, 
and the model function fitted to the observation is shown with a solid line.
The error bars represent one sigma statistical errors.
The horizontal dashed line in figures~\ref{fig:density_dim} and \ref{fig:density_bright}
represents the fitting parameter $<n_{\mbox{\scriptsize bg}}>$ of equation~(\ref{eq:model}).
The bin size was fixed to 0.1~Mpc for the redshift groups of z1 and z2, and it was fixed
to 0.2~Mpc starting from 0.1~Mpc for the z3, z4 and z5 groups.
For the z3, z4 and z5 groups, the data of $<$ 0.1~Mpc was ignored, since a large fraction 
of this distance range was masked to avoid the effect of AGN light.

In the right panels of the figures, the corresponding projected correlation functions 
$\omega(r_{p})$ are also shown.
We fixed the slope parameter $\gamma$ to 1.8, which is a canonical value 
obtained from auto-correlation analysis among typical local galaxies by many other
works (e.g.  \cite{Zehavi_2005}) and has also been obtained for QSO auto-correlation 
and QSO-galaxy cross-correlation studies (e.g. 1.9 for \cite{Ross_2009} and 1.5--1.9 
for \cite{Coil_2007}).
To verify the choice of value for freezing $\gamma$, we performed fitting to a larger
sample obtained by combining z1-D, z2-D and z2-B samples by allowing $\gamma$ to float.
We obtained $\gamma = 1.80 \pm 0.13$ for the sample, thus the choice of 1.8 for 
the parameter $\gamma$ is reasonable in this analysis.
%%
%% $<\rho_{0}>$ is an average of $\rho_{i,0}$ for each AGN in the group,
%% and it is fixed to the value calculated as described in section~\ref{sec:rho0}.
%%

%%
%%
%%
\section{Results}
\label{sec:result}

\subsection{Overview of the results and comparison with other experiments}

Evidence of galaxy clustering around the AGNs was detected with a 90\%
confidence level (C.L.) for all the groups except for the z5-B group
The fitting parameters of the model function
are summarized in table~\ref{tbl:fit_param}.
We obtained cross-correlation lengths of
$4.7^{+1.3}_{-0.7}$, 
$5.8^{+1.9}_{-1.1}$ and
$7.6^{+3.2}_{-1.9}$~$h^{-1}$Mpc for the dim AGN groups of z1-D, z2-D, and z3-D,
respectively.
For the bright AGN groups, cross-correlation lengths of
$6.3^{+2.4}_{-1.5}$,
$5.1^{+2.7}_{-1.5}$,
$11.1^{+6.1}_{-2.7}$~$h^{-1}$~Mpc were obtained for the z2-B, z3-B, and z4-B group,
respectively.
The errors quoted here indicate the sum of a systematic error estimated from 
uncertainty of $\rho_{0}$ and a statistical error of one sigma. 
The lower (upper) systematic error was given by a difference of
$r_{0}$ values obtained by fixing $<\rho_{0}>$ of the model function to
$<\rho_{\mbox{\scriptsize 0,mid}}>$ and $<\rho_{\mbox{\scriptsize 0,upp}}>$ 
($<\rho_{\mbox{\scriptsize 0,low}}>$), where 
$<\rho_{\mbox{\scriptsize 0,mid}}>$, 
$<\rho_{\mbox{\scriptsize 0,upp}}>$, and
$<\rho_{\mbox{\scriptsize 0,low}}>$ are 
the average densities derived for $F_{\mbox{\scriptsize limit}} =$ 0.6, 0.8, and 0.1, respectively.
We ignored covariances among the number density bins.
The effect of the covariance to the error estimate for $r_{0}$ will be discussed 
in section~\ref{sec:systematic}.
For the z5-B AGN group, no significant clustering 
signature was observed, and an upper limit of 13.1~$h^{-1}$~Mpc was obtained.
This upper limit includes systematic error due to uncertainty of $\rho_{0}$ and
statistical error of one sigma.

%%
%% r0 vs Redshift の図
%%
In figure~\ref{fig:r0_z}, the cross-correlation length obtained for each AGN
group is plotted as a function of redshift. 
The bright AGN groups (z2-B, z3-B, z4-B) are shown as closed circles,
and the dim AGN groups (z1-D, z2-D, z3-D) are shown as closed triangles.
The upper limit is shown for the z5-B group.
For comparisons, results of
AGN-galaxy cross-correlation analysis by other authors 
(\cite{Bornancini_2007,Coil_2007,Norman_2009,Hickox_2009})
are also shown in 
the figure.
The auto-correlation length of galaxy or QSO obtained by 
other authors 
(\cite{Ma_2009,Zehavi_2005,Hawkins_2003,Wake_2008,Ross_2009,Shen_2007})
are also shown in the same figure.

It may not be appropriate to compare the clustering strength only with the
correlation length, since it depends on the assumed or fitted slope
parameter $\gamma$.
The clustering strength is often compared in $\sigma_{8}$, which
is the root mean square of correlation function 
in the sphere with a comoving radius of $r_{\mbox{\scriptsize max}} = $ 8 $h^{-1}$Mpc, and 
calculated as (\cite{Miyaji_2007, Peebles_1980}):
\begin{equation}
   \sigma_{\mbox{\scriptsize 8}}^{2} % = \int \int \xi(|{\bf r_{1}-r_{2}}|) dV_{1} dV_{2}/V^{2} 
                     = (r_{0}/r_{\mbox{\scriptsize max}})^{\gamma} J_{2},
\end{equation}
\begin{equation}
   J_{2} = 72 / [ (3-\gamma) (4-\gamma) (6-\gamma) 2^{\gamma} ].
\end{equation}
%%
%% In the case of AGN-galaxy correlation, this parameter represents the rms
%% excess density of galaxies within a comoving distance of 8 h$^{-1}$Mpc.
%%
Using these formulae, we converted a correlation length of each experiment to
$\sigma_{8}$, and they are compared in figure~\ref{fig:sigma8_z}.

\subsection{Results for the low redshift ($z <$ 0.9) AGN groups}

\begin{figure*}
  \begin{center}
    \FigureFile(\textwidth,80mm){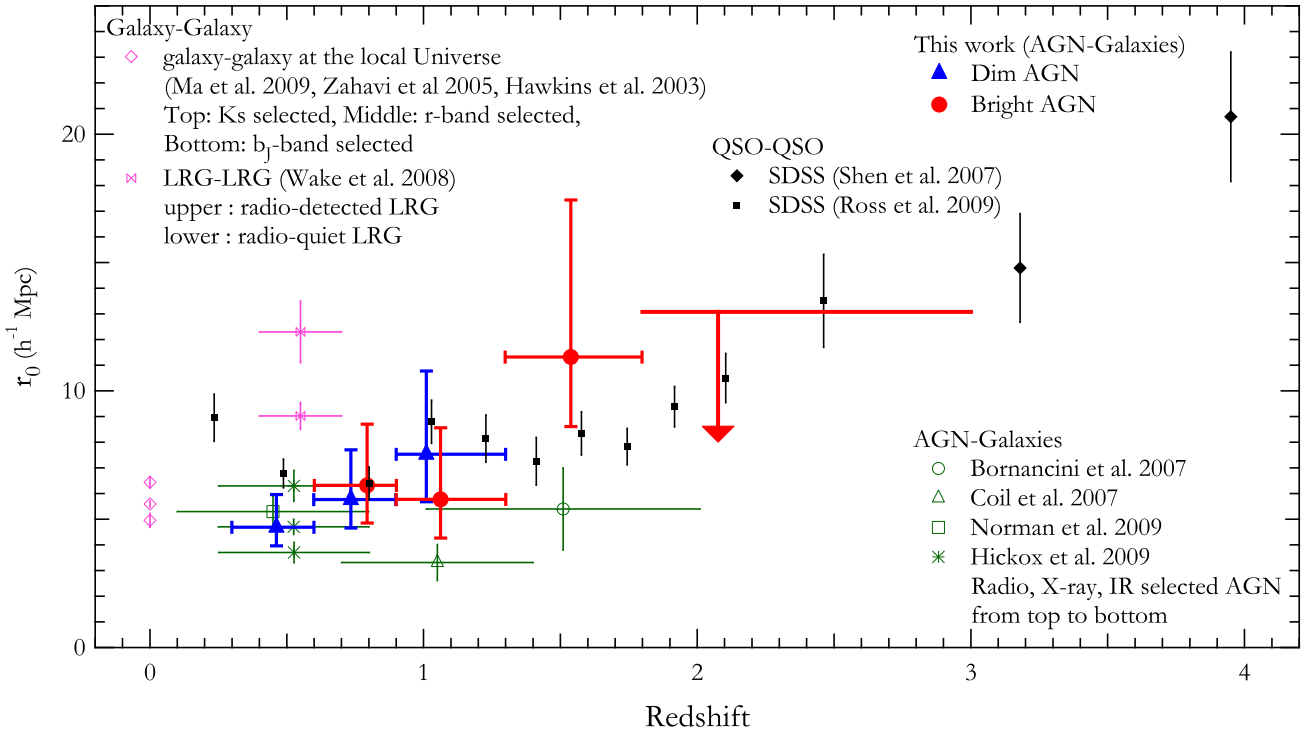}
  \end{center}
  \caption{Measured correlation length, $r_{0}$, as a function of redshift.
The error bars of our results indicate sum of the systematic error estimated from 
uncertainty of $\rho_{0}$ and the statistical error of one sigma. 
Covariances among the number density bins are not taken into account to derive
the statistical error.
$\sigma_{\mbox{\scriptsize Jackknife}}$ of table~\ref{tbl:sigma_poisson_jackknife} gives an upper 
limit on the statistical error which includes covariance.
The closed triangles represent the result for dim AGN groups, and 
the closed circles represent the result for bright AGN groups. 
The data points of this work is the result derived for combined UKIDSS and
Suprime-Cam samples.
The result for z5-B are shown with the upper limits.
It should be noted that each data point is derived from different types of AGN
and/or galaxy sample, thus this figure is for comparative purposes only.
}
  \label{fig:r0_z}
\end{figure*}

\begin{figure*}
  \begin{center}
    \FigureFile(\textwidth,80mm){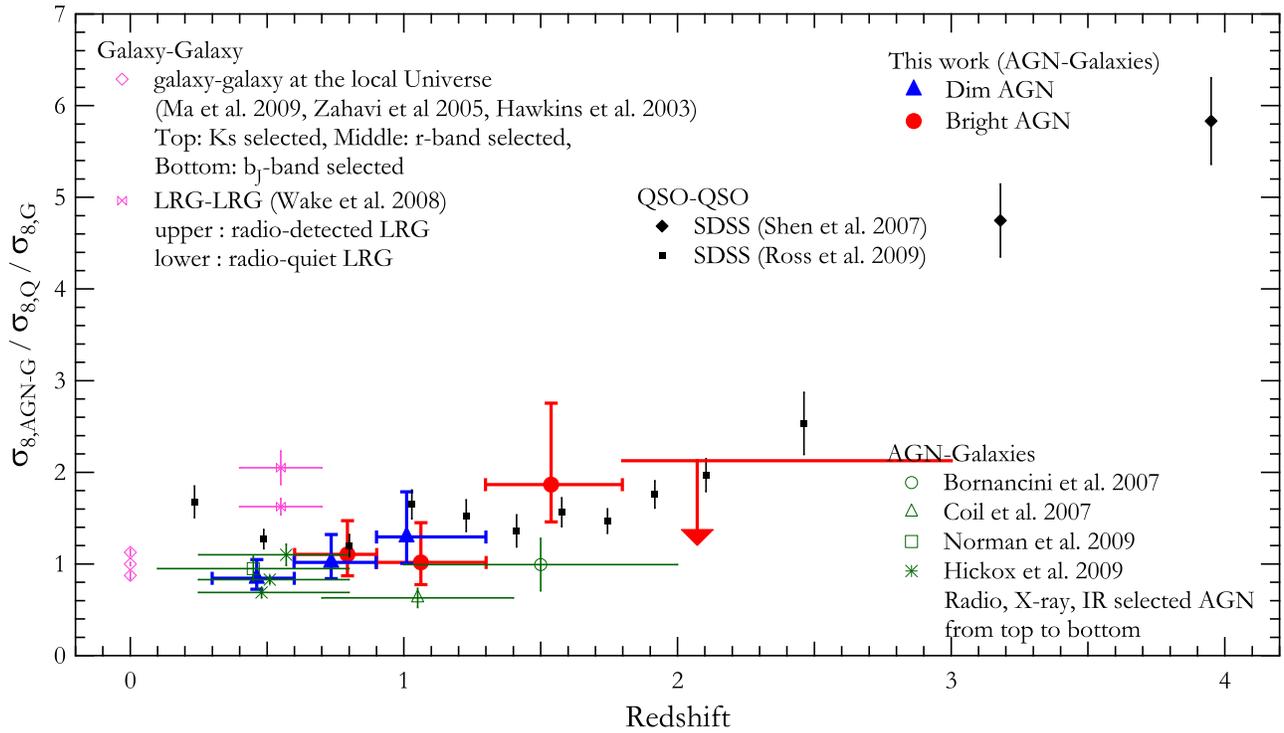}
  \end{center}
  \caption{The $\sigma_{8}$ of the galaxies around AGNs ($\sigma_{\mbox{\scriptsize 8,AGN-G}}$),
QSOs ($\sigma_{\mbox{\scriptsize 8,Q}}$), and galaxies ($\sigma_{\mbox{\scriptsize 8,G}}$) are 
shown as a function of redshift. The error bars of our results indicate sum of the
systematic error estimated from uncertainty of $\rho_{0}$ and the statistical error of 
one sigma.
Covariances among the number density bins are not taken into account to derive
the statistical error.
See subsection~\ref{sec:systematic} for a discussion on the covariance.
The data points of this work is the result derived for combined UKIDSS and
Suprime-Cam samples.
It should be noted that each data point is derived from different types of AGN
and/or galaxy sample, thus this figure is for comparative purposes only.
}
  \label{fig:sigma8_z}
\end{figure*}

%%
%% z1 & z2 の結果
%%
The results for the low redshift groups ($z=$0.3--0.9, z1 and z2) are consistent 
with the existing measurements.
\citet{Norman_2009} measured the projected cross-correlation between 420 QSOs 
and 4,975 luminous red galaxies (LRG) at redshifts from 0.2 to 0.8 (an open
square in figures~\ref{fig:r0_z} and \ref{fig:sigma8_z}) based on the data of 2QZ and 2SLAQ survey.
Our measurements of $\sigma_{8}$ at $z=$ 0.3--0.9 (z1-D, z2-D, z3-B) agree 
with their values, although our galaxy sample is a mixture of dim galaxies
detected with the Suprime-Cam and bright galaxies detected with the UKIDSS survey.
\citet{Hickox_2009} also measured the cross-correlation of three types
of AGNs and galaxies (asterisks in figures~\ref{fig:r0_z}  and \ref{fig:sigma8_z}; from top to bottom
radio, X-ray and IR selected AGNs) based on the AGN and Galaxy Evolution 
Survey (AGES) and Bootes multi-wavelength survey.
Their sample contains 598 AGNs at redshift of 0.25 -- 0.8.
Our results are nearly consistent with their result for all the three types within
the errors, but closest to the X-ray AGN result.
The clustering of local galaxies measured by \citet{Ma_2009} 
(K-band selected galaxy), \citet{Zehavi_2005} (r-band selected galaxy), and 
\citet{Hawkins_2003} (b$_{\mbox{\scriptsize J}}$-band selected galaxy) are 5--7~Mpc (open diamonds
in figure~\ref{fig:r0_z} from top to bottom, respectively).
Our values for AGN-galaxy cross-correlation length at $z=$ 0.3--0.9 are, therefore, 
almost the same as the auto-correlation length of the local galaxies.

\subsection{Galaxy selection effect for the low redshift groups}

%%
%% Our galaxy sample is a mixture of galaxies detected by UKIDSS and Suprime-Cam.
%%
From the observational facts that at the local Universe red early-type galaxies 
are more clustered than blue late-type galaxies and bright galaxies are more 
clustered than dim galaxies, we can expect different clustering
strength for UKIDSS and Suprime-Cam galaxies samples.
To see the difference, we calculated the AGN-galaxy cross correlation function 
for UKIDSS selected galaxies and Suprime-Cam selected galaxies independently
for the z1-D and z2-D AGN groups.
The results are summarized in table~\ref{tbl:fit_param} and labeled 
as ``OPT'' for the Suprime-Cam galaxy samples and ``IR'' for the UKIDSS galaxy 
samples.

\begin{figure*}
  \begin{center}
    \FigureFile(0.7\textwidth,80mm){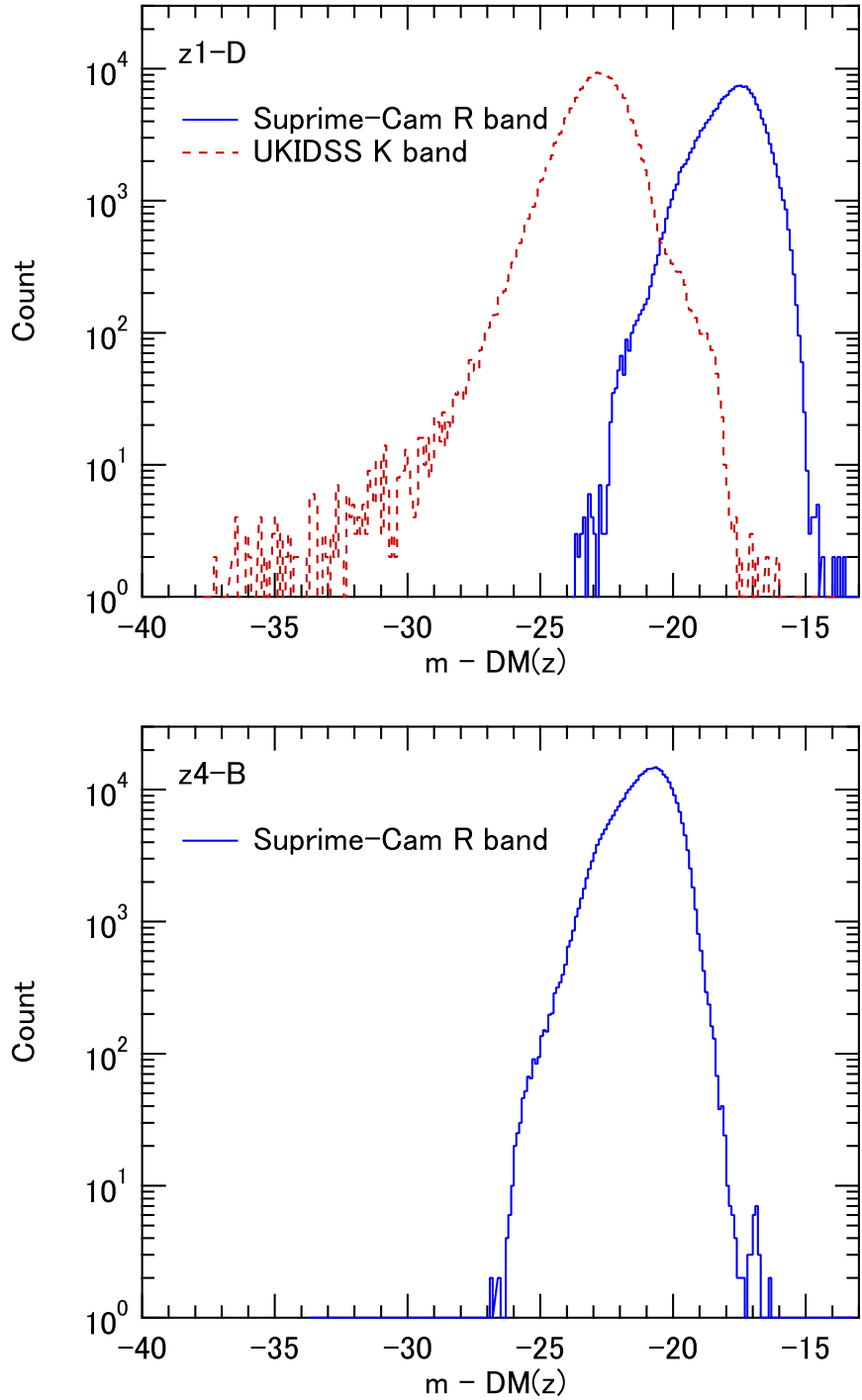}
  \end{center}
  \caption{
The distribution of $m - DM(z)$, where $m$ is
an apparent magnitude of detected objects, and $DM(z)$ is a distance modulus
corresponding to the AGN redshift.
The top panel shows the distribution for z1-D galaxy samples observed with Suprime-Cam in R band 
(solid line) and in UKIDSS K band (dashed line).
The bottom panel shows the distribution for z4-B galaxy samples observed with Suprime-Cam in R
band.
}
  \label{fig:maghist}
\end{figure*}

For the z1-D AGN group, larger correlation length was obtained for
UKIDSS galaxies samples ($r_{0} = 6.8^{+1.9}_{-1.0}$~$h^{-1}$Mpc ) than that for 
Suprime-Cam galaxies samples ($r_{0} = 3.0^{+1.5}_{-1.1}$~$h^{-1}$Mpc).
The clustering of UKIDSS galaxies around z1-D AGNs is consistent with
that of K-band selected galaxies measured by \citet{Ma_2009} at the 
local Universe.
As for the Suprime-Cam galaxies, the clustering is as small as 
that of dim galaxies ($M_{r} \ge -19$) measured by \citet{Zehavi_2005}.
At this redshift range (z1), galaxies in the Suprime-Cam sample are dimmer
than $M \sim -20$ 
as the brighter galaxies were masked before the source 
extraction, while the UKIDSS galaxy sample consists of bright galaxies 
($M < -21$) (see top panel of figure~\ref{fig:maghist}).
Thus the Suprime-Cam galaxies sample is strongly biased to dim 
galaxies (expected to be in the range of $M = -16$ to $-20$), while 
the UKIDSS sample is biased to bright galaxies ($M < -21$).
The cross-correlation length obtained for the Suprime-Cam galaxies can be 
consistent with the auto-correlation of optically selected dim galaxies at
the local Universe, if the luminosity dependence measured by \citet{Zehavi_2005}
is taken into account.
%%
%% The correlation length obtained for the UKIDSS galaxies is also consistent
%% with the result for bright galaxy samples obtained by \citet{Zehavi_2005}.
%%

As for the z2-D group, cross-correlation length of $7.4^{+4.7}_{-2.1}$
and $5.0^{+1.6}_{-1.0}$~$h^{-1}$Mpc were obtained
for the UKIDSS and Suprime-Cam samples, respectively.
The difference was smaller than that of the z1-D group, and they were nearly
consistent within their uncertainties.
At this redshift range, maximum brightness of the Suprime-Cam galaxies
comes down to $M \sim -22$, which almost equals or is
brighter than the characteristic luminosity $L_{*}$ (break point of a luminosity 
function).
Thus, it is expected that the clustering of Suprime-Cam galaxies around the AGNs
is enhanced due to the increase in the fraction of bright galaxies in the sample.
We conclude that the clustering of UKIDSS and Suprime-Cam galaxies around 
AGNs can be explained by the luminosity dependence of
galaxy clustering as observed in the local Universe.

The luminosity range of galaxies in the samples of \citet{Norman_2009} 
(i $\ge$ 19.8) and \citet{Hickox_2009} (I $\ge$20) are almost the same 
as that of the UKIDSS LAS data.
$\sigma_{8,\mbox{\scriptsize AGN-G}}$ calculated for the UKIDSS samples are 1.18$^{+0.29}_{-0.16}$ and
1.27$^{+0.71}_{-0.33}$ for the z1-D and z2-D AGN group respectively, and they are
consistent with the result of \citet{Norman_2009} and the result for 
radio-selected AGN sample of \citet{Hickox_2009}.

\subsection{Results for the high redshift ($z \ge$ 0.9) AGN groups}

%%
%% z = 0.9 -- 1.3 の結果
%%
The result for the z3-D group is significantly larger than the result 
by \citet{Coil_2007}.
They measured quasar-galaxy cross-correlation length 
from 52 QSOs at redshifts from 0.7 to 1.4 based on the SDSS and DEEP2 survey 
(an open triangle in figures~\ref{fig:r0_z} and \ref{fig:sigma8_z}).
The range of absolute magnitude of AGNs used in \citet{Coil_2007} is 
from $-$26 to $-$20, which is almost equivalent to our dim AGN group.
\citet{Coil_2007} measured the correlation length using the data at
a distance scale from 0.1 to 10~Mpc.
They used $\gamma$ as a free parameter, then obtained 1.55 for $\gamma$.
Thus we fit the model function to our data with a fixed $\gamma$ of 1.55.
Then we obtained much longer correlation length, which
is not consistent with the result by \citet{Coil_2007}.
%%
%% The difference may come from the difference of galaxy samples used.
%%
%% The DEEP2 galaxies are brighter than the Suprime-Cam samples,
%% thus the difference may be attributed to the difference in the clustering 
%% properties of a dimmer-blue galaxy and a bright-red galaxy.
%%
It is possible that the difference is due to the cosmic variance, as
the number of well separated fields is only four for the DEEP2 dataset.

It should be noted that our clustering measurement is dominated by
dim galaxies near the detection threshold, while
the other experiment conducting spectroscopic observations is mostly 
targeted to the bright end of the luminosity function.
Therefore, our measurement has sensitivity to the clustering 
of relatively dim galaxies which are not measured by spectroscopic 
observations.
In addition to that, our optical measurement is biased to blue galaxy at 
this redshift range, thus it is expected that star-forming galaxies are 
dominant in our galaxy sample.
In the case of spectroscopic observation, a red galaxy is preferentially 
selected as a target since the redshift estimate for a
red galaxy is more robust than that for a blue galaxy owning to the existence
of a sharp 4,000 \mbox{\AA} break in the spectrum.
The $\sigma_{\mbox{\scriptsize 8,AGN-G}}$ obtained for the z3-D group is almost comparable 
with $\sigma_{\mbox{\scriptsize 8,Q}}$ derived from the QSO auto-correlation length 
by \citet{Ross_2009}, and with $\sigma_{\mbox{\scriptsize 8,G}}$ derived from the 
auto-correlation length of radio-quiet Luminous Red Galaxies (LRG) at lower 
redshift by \citet{Wake_2008}.
The cross-correlation length for the bright AGN group at the same redshift (z3-B) 
is consistent with that for the z3-D group, and no significant evidence of
AGN luminosity dependence is found.
\citet{Shen_2009} examined luminosity dependence of the clustering of
SDSS QSOs.
They also found no significant evidence for the luminosity dependence, which
is consistent with our observation.

%% 
%% z4 の結果
%%
The distribution of galaxy number density for the bright AGN group at redshift 
1.3--1.8 (z4-B) shows relatively flat excess at $<$ 2.5~Mpc.
Significance of the excess over the power-law model function at $r_{p}$ = 1.3 to 
2.7~Mpc is 3.4~$\sigma$.
%%
%% $\chi^{2}$ relative to the model function at $r_{p} < 3~Mpc$ is 22.24 for 15 dof, 
%% which corresponds to 10\% of chance probability.
%%
This feature may indicate that a significant number of the AGNs are not located 
at the center of a galaxy group but distributed over a scale of 5~Mpc.
The clustering strength, $\sigma_{\mbox{\scriptsize 8,AGN-G}}$, obtained for z4-B group 
is nearly consistent within the quoted error with the QSO clustering 
strength $\sigma_{\mbox{\scriptsize 8,Q}}$ 
calculated from the observational result by \cite{Ross_2009} at the same redshift.

\begin{figure*}
  \begin{center}
    \FigureFile(1.0\textwidth,80mm){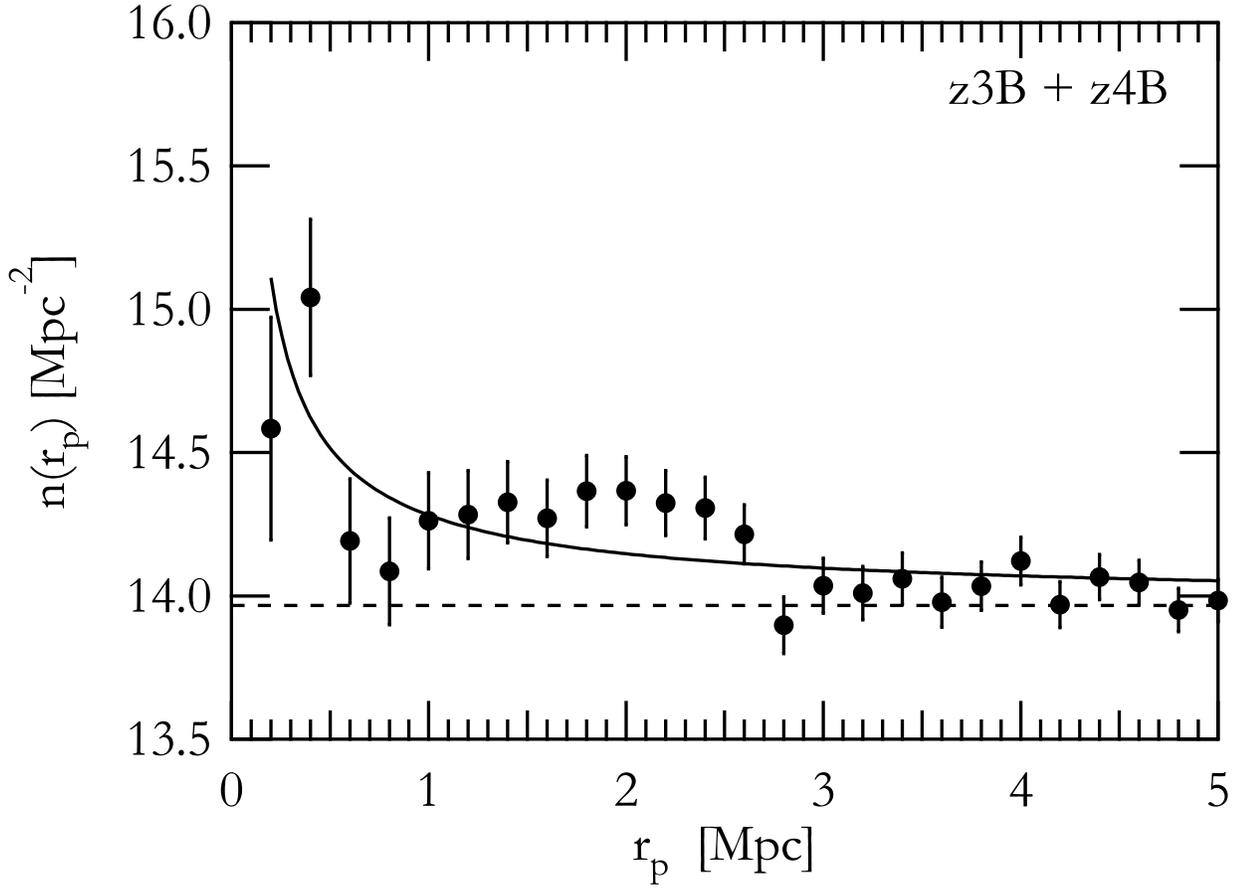}
  \end{center}
  \caption{
Averaged number density around z3-B and z4-B AGNs.
}
  \label{fig:density_z34B}
\end{figure*}

\citet{Bornancini_2007} have examined the cross-correlation between the 13 SDSS QSOs 
and the distant red galaxies (DRG; $J-K_{s} > 2.3$) at redshifts from 1 to 2 
(an open circle in figures~\ref{fig:r0_z} and \ref{fig:sigma8_z}). 
Their AGN sample has a luminosity range similar to our bright AGN group.
They reported a cross-correlation length of 5.4$\pm$1.6~$h^{-1}$Mpc.
In order to compare to the results for this sample, for which the redshift range
extends over those covered by our z3 and z4 groups, 
we calculated a correlation length and $\sigma_{\mbox{\scriptsize 8,AGN-G}}$
for a combined group of z3-B and z4-B to be 7.5$^{+3.3}_{-1.6}$~$h^{-1}$Mpc 
and 1.3$^{+0.5}_{-0.3}$, respectively.
Although this result is almost consistent with the value of \citet{Bornancini_2007} 
within the quoted error, the distribution of number density shows a flat excess
at 1--2.5~Mpc (figure~\ref{fig:density_z34B}), which was significant for z4-B 
group and barely seen also in the z3-B group.
This feature was not observed in the data of \citet{Bornancini_2007}.
The magnitude range of the galaxy sample used by \citet{Bornancini_2007} is
$K \le 19.5 (\mbox{Vega}) \sim 21.4 (\mbox{AB})$, while our galaxy sample 
of Suprime-Cam is dimmer than m$_{\mbox{\scriptsize AB}}$ $\sim$ 22 on average.
Thus the difference in the number density profile might be due to the difference in
brightness and/or observation band of the galaxy samples.
%%

%%
%% z5 の結果
%%
For the AGN group at redshift 1.8--3.0 (z5-B), we could not find any evidence of 
a clustering signature.
At this high redshift range, the number of galaxies which are brighter than the
limiting magnitude is very poor, and the clustering signal easily disappears
due to the foreground galaxies.

\subsection{Galaxy selection effect for the high redshift groups}

\begin{figure*}
  \begin{center}
    \FigureFile(1.0\textwidth,80mm){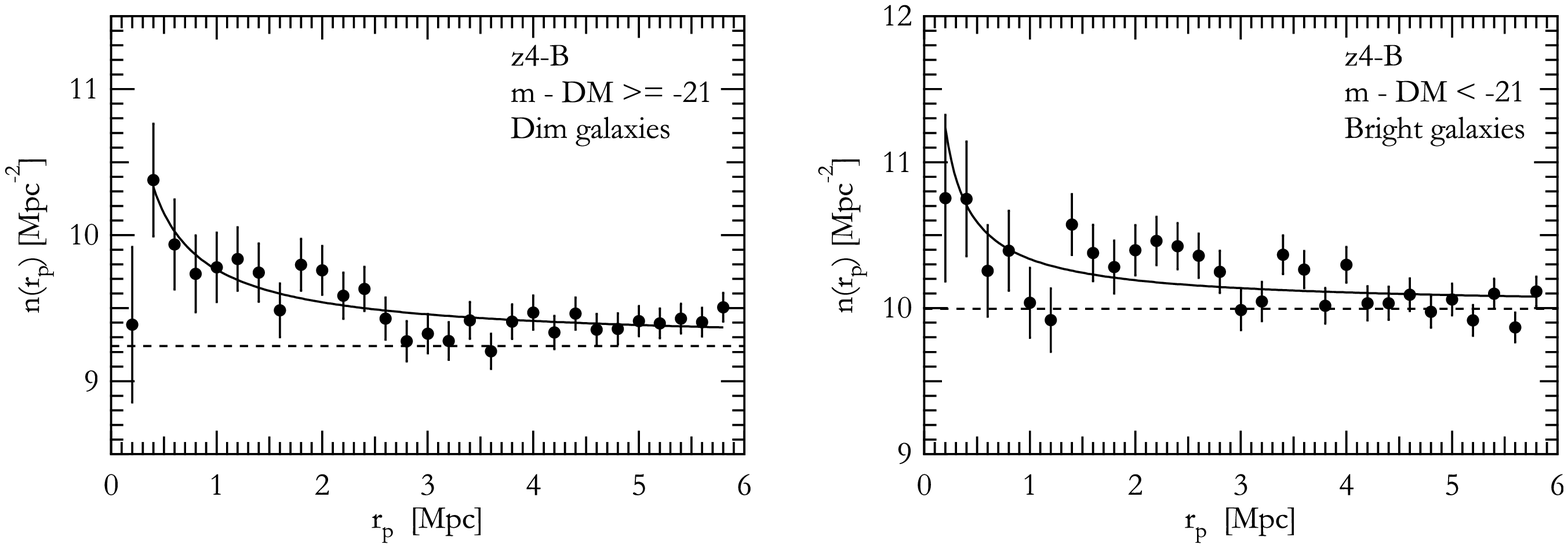}
  \end{center}
  \caption{
Averaged number density of dim sources ($m \ge -21 + DM$) (left panels) 
and of bright sources ($m < -21 + DM$) (right panels)
around z4-B AGNs, where DM
represents a distance modulus corresponding to AGN redshift.
The solid lines represent model function of equation~(\ref{eq:model}) fitted to the
observation data. The fit was done for distance range of 0.3--6.0 Mpc for the case of
dim sources, while 0.1--6.0 Mpc for bright sources. The horizontal lines represent
the fitting parameter $n_{\mbox{\scriptsize bg}}$.
}
  \label{fig:density_M21}
\end{figure*}

%%
%% 銀河の明るさ依存性について
%% 
The detection limit of galaxies for a lower redshift group is dimmer 
in absolute magnitude than that for a higher redshift group.
From observations of the local Universe, it is known that brighter galaxies 
usually have stronger clustering.
\cite{Zehavi_2005} analyzed data from SDSS DR2 and derived the luminosity
dependence of the galaxy clustering.
They showed that the galaxy auto-correlation function increases continuously
with luminosity, and above the characteristic luminosity $L_{*}$ ($M_{r} \sim -20.5)$
it increases more rapidly.
The measured auto-correlation lengths were
9.81$\pm$ 0.39~$h^{-1}$Mpc for $M_{r} < -22$, 
decreased rapidly down to 6.16$\pm$0.17 for $-22 < M_{r} < -21$,
and decreased rather slowly down to 2.68$\pm$0.39 for $-18 < M_{r} < -17$.
Thus it is expected that the observed clustering of galaxies around AGNs 
can be stronger at higher redshift due to the bias to brighter galaxies.
In figure~\ref{fig:density_M21}, the distributions of averaged number 
density of bright ($m < -21 + DM$, left panel) and 
dim ($m \ge -21 + DM$, right panel) 
galaxies around the AGNs are shown for the z4-B AGN group.
The range of absolute magnitude for both samples can be inferred from 
the bottom panel of figure~\ref{fig:maghist}, which shows the distribution
of $m - DM(z)$ for R-band observations.
They are expected to be in the range of $-24$ to $-21$  and
$-21$ to $-19$ for the bright and dim galaxy samples, respectively.
As can be seen from the figure~\ref{fig:density_M21}, the excess number 
densities observed for dim and bright galaxy samples are almost comparable
to each other.
The excess of source count at $r_{p}<2.5$Mpc is 9.3$\pm$1.2 and 7.0$\pm$1.2
for the dim and bright galaxy sample, respectively.
Thus half of the clustering found for the z4-B group is contributed from bright
galaxies with absolute magnitude less than $-21$.
It should be noted that this does not necessarily mean that the clustering
strength of dim galaxy sample and bright galaxy sample is equivalent, because
the average numbers of detectable galaxies at the AGN redshift are different for
the two samples.
Since $\rho_{0}$ for each of the bright and dim galaxy sample cannot be determined
by the method as shown in section~\ref{sec:rho0}, we estimated the upper limit for
the number density of bright galaxies ($M < -21$) 
by taking an average of smaller values of the followings calculated for each AGN sample: 
$\rho_{0}$ calculated for the whole galaxies as described in section~\ref{sec:rho0} or
integral of the luminosity function up to $M = -21$ calculated in the same way as
the calculation of $\rho_{0}$ for whole galaxies.
Then we obtained the upper limit of the average number density for the bright 
galaxies of the z4-B group as $\sim 0.26 \times 10^{-3}$ Mpc$^{-2}$.
The average density estimated for the whole galaxies sample is 
$1.3 \times 10^{-3}$~Mpc$^{-2}$, thus the number density of the bright galaxies is less 
than 1/4 of the dim galaxies ($M \ge -21$).
The lower limit of correlation length for the bright galaxy sample was calculated
as 18~$h^{-1}$Mpc by adapting $\rho_{0} = 0.26 \times 10^{-3}$ Mpc$^{-2}$ in the model fitting.
This correlation length is significantly larger than that measured for a bright galaxy
at the local Universe.
%%
%% Thus, it is expected that the excess number density for the bright galaxy sample is
%% three times smaller than that for the dim galaxy sample, if it has the same clustering
%% strength as the dim galaxy.
%%
%% The observed excess is consistent with this expectation within the statistical 
%% uncertainty.
%%
Therefore the large correlation length found for the z4-B group cannot be
explained simply by the luminosity dependence observed in the local Universe.
%%

%%
%% 銀河種別に対するバイアス
%%
There would be another possibility that the redshift dependence is caused by
selection bias to a type of galaxy, that is, red or blue galaxy.
It is known that the red early type galaxies are clustered stronger than 
the blue late type galaxies at the local Universe.
Although we used UKIDSS IR data, the observations were too shallow 
to detect the high redshift galaxies and they were almost not used for the
high redshift groups.
Thus our high redshift dataset is biased to blue galaxies.
This is contrary to the expectation that the larger correlation length for the z4-B 
group is due to the selection bias to the redder galaxies which are strongly 
clustered in the local Universe.

\subsection{Examination for other possible systematic effects}
\label{sec:systematic}
To examine the possibility that the high clustering observed for the z4-B
group was due to accidental alignment of high density regions,
we analyzed eight independent offset fields of the z4-B AGNs.
The offset fields were taken apart from each AGN by 10~Mpc in the 
directions of right ascension and/or declination, so that the offset fields
did not overlap with the high density region around AGN.
We applied the same analysis procedure to these offset fields as was applied to 
the AGN fields.
%%
%% The right bottom panel of the Figure~\ref{fig:density_x6} shows the averaged
%% density distribution around the offset positions, which was obtained by taking 
%% an average over one of the eight offset fields that meet the criteria 
%% adopted to the AGN dataset.
%%
%% No systematic excess was seen for the offset field.
%%
We derived density distributions averaged over each of the eight 
offset fields.
The significance of excess or deficit at $r=$0--2.5~Mpc to the average
density at $r=$4--6~Mpc were less than two sigmas,
while the significance for the z4-B AGNs was 5.8~$\sigma$.
As no significant systematic excess was seen for the offset fields,
we conclude that the excess seen for the AGN dataset is most likely
due to galaxies associated with the AGNs.

\begin{figure*}
  \begin{center}
    \FigureFile(1.0\textwidth,80mm){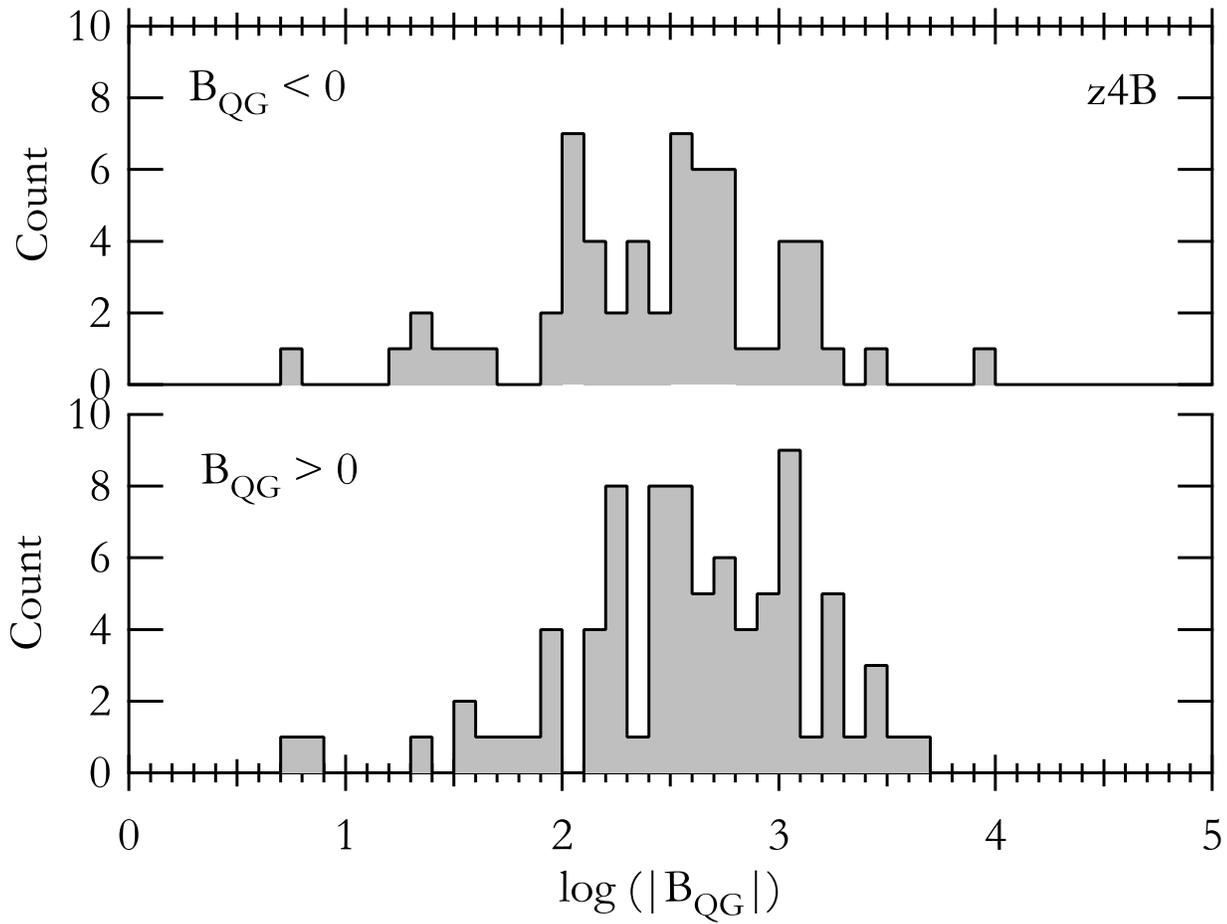}
  \end{center}
  \caption{The logarithmic distribution of clustering coefficient $B_{QG}$ for 142 z4-B samples.
The top panel is for negative $B_{QG}$ and the bottom panel is for positive $B_{QG}$.}
  \label{fig:Bqg_z4B}
\end{figure*}

\begin{figure*}
  \begin{center}
    \FigureFile(1.0\textwidth,80mm){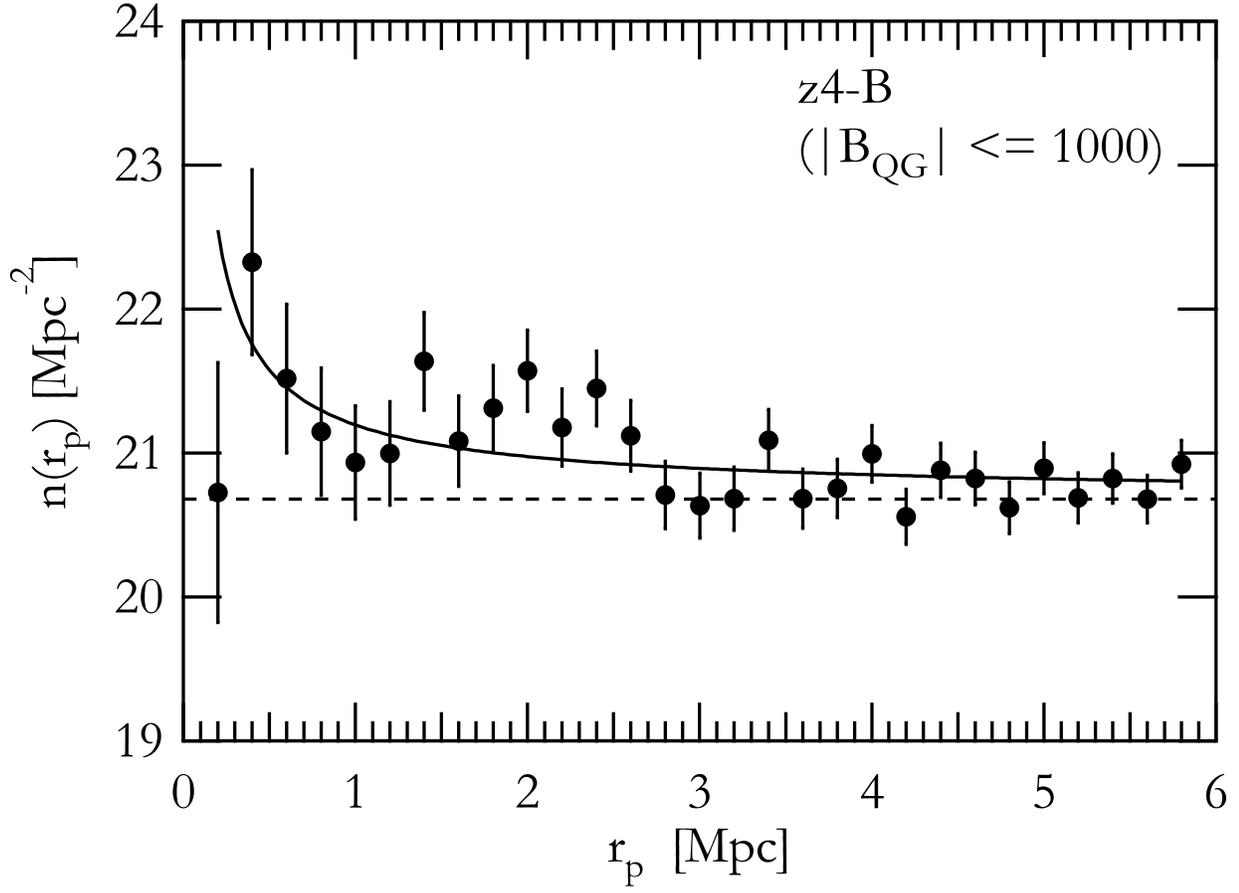}
  \end{center}
  \caption{
Averaged number density around z4-B AGNs which satisfy the condition
$|B_{\mbox{\scriptsize QG}}| \le 1000$.
}
  \label{fig:density_z4B_Bqg3}
\end{figure*}

We also examined the possibility that the excess was caused by only a single or a
few samples which had large clustering.
We calculated the average number density around z4-B AGNs for which absolute
value of $B_{\mbox{\scriptsize QG}}$ was less than or equal to 1,000 rather than our usually
adapted limit of $10^{4}$.
The $B_{\mbox{\scriptsize QG}}$ distribution for the z4-B AGN groups is shown in 
figure~\ref{fig:Bqg_z4B}, and 110 out of the 142 AGN samples satisfied 
this condition.
The number density distribution is shown in figure~\ref{fig:density_z4B_Bqg3}.
The excess at $r<2.5$ Mpc is clearly found also for this relatively small clustering
samples of $|B_{\mbox{\scriptsize QG}}| \le 1,000$.
The calculated cross-correlation length was 8.7$^{+5.2}_{-2.5}$~$h^{-1}$Mpc.
From this result, we conclude that the excess seen in the z4-B group is 
not created just from a single or a few samples with the largest clustering, but
from a wide range of samples of the z4-B group.

\begin{figure*}
  \begin{center}
    \FigureFile(1.0\textwidth,80mm){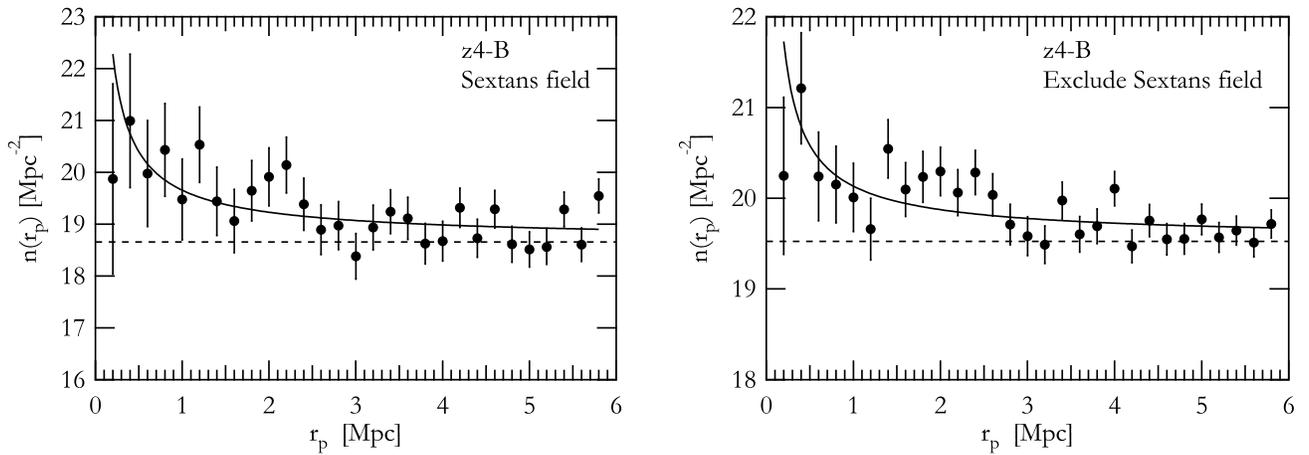}
  \end{center}
  \caption{
Averaged number density around z4-B AGNs not located in the Sextans field (left) 
and AGNs located in the Sextans field (right).
}
  \label{fig:density_sextans}
\end{figure*}

As described in section~\ref{sec:dataset}, 20\% of the z4-B AGN samples are located
in the Sextans field of RA $=$ 150$^{\circ}$ -- 157$^{\circ}$ and Dec $= -4^{\circ}$ --
$+1^{\circ}$.
To see the effect of the concentration to this specific field, we derived the
distribution of number density for z4-B AGNs excluding the samples located in the
Sextans field, and showed it in the left panel of figure~\ref{fig:density_sextans}.
In the right panel of the same figure, the density distribution for z4-B AGNs located
in the Sextans field is also shown.
No significant difference is seen for the two density profiles.
The cross-correlation length for the z4-B samples excluding the AGNs of Sextans filed is
10.0$^{+2.6}_{-2.6}$~$h^{-1}$Mpc, and it is consistent with the result for the whole z4-B
samples.
Thus, we can conclude that the result for the z4-B group is not strongly biased by
the samples of the Sextans field.

\begin{figure*}
  \begin{center}
    \FigureFile(1.0\textwidth,80mm){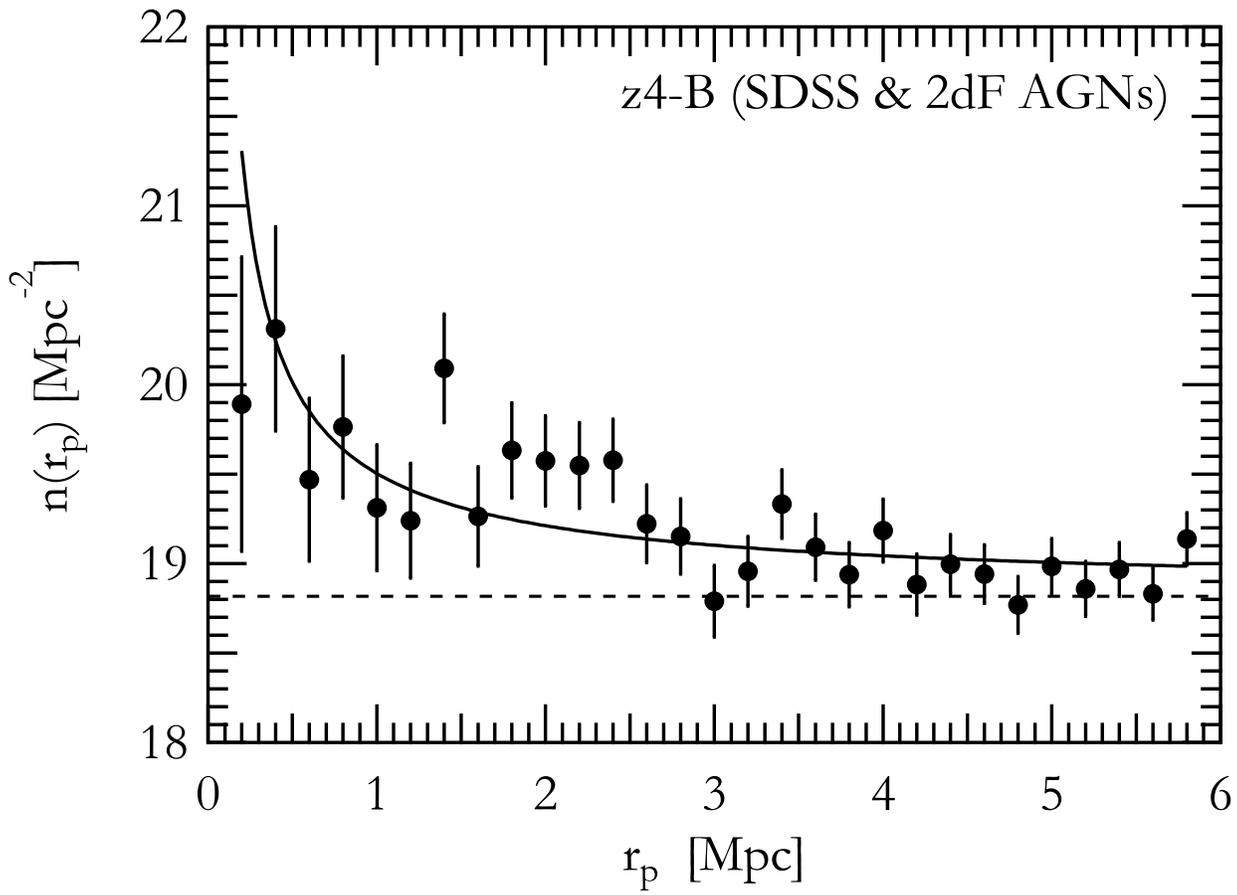}
  \end{center}
  \caption{
Averaged number density around z4-B AGNs selected from SDSS and 2dF QSO catalogs.
}
  \label{fig:density_sdss_2df}
\end{figure*}

Our AGN sample consists of various types of AGNs.
To examine the effect of this heterogeneous sample of AGNs,
we performed the calculation for AGNs which come from SDSS 
and 2dF QSO catalogs.
The number density distribution for this homogeneous samples 
is shown in figure~\ref{fig:density_sdss_2df}.
The density distribution obtained for the AGNs which come
from homogeneous optical selection is almost identical
to that for the heterogeneous sample.
The calculated correlation length for the homogeneous sample
is $11.7^{+6.6}_{-2.9}$~$h^{-1}$Mpc,
which is consistent with the result for the heterogeneous sample.
Thus we can conclude that the large clustering observed for the
z4-B group is not due to the heterogeneity of the sample.
In driving the errors of cross-correlation length $r_{0}$, we ignored 
covariance among the galaxy number density at each distance bin.
The densities in different distance bins are not independent from one another, thus
the covariances have to be taken into account for error estimate of $r_{0}$.
To investigate the effect of the covariance, the error of $r_{0}$ 
was estimated from the jackknife method (\cite{Norberg_2009}) and
compared with the error estimated only from Poisson statistic 
ignoring the covariance.
The jackknife resamplings were made by omitting, in turn, each of 
the AGN samples.
We found that the covariance estimated from the jackknife method
defined in \citet{Norberg_2009} were significantly overestimated 
owing to the large fluctuation of $n_{\mbox{\scriptsize bg}}$ for each AGN sample.
Thus we decided to calculate $r_{0}$ for each jackknife resampling,
and estimate the error of $r_{0}$ from their variance.
The error of $r_{0}$, which is denoted as $\sigma_{\mbox{\scriptsize Jackknife}}$,
was estimated from the following equation:
\begin{equation}
   \sigma_{\mbox{\scriptsize Jackknife}}^{2} = \frac{N-1}{N} 
     \sum_{i=1}^{N} (r_{0,i} - <r_{0}>)^{2},
\end{equation}
where $r_{0,i}$ is the cross-correlation length obtained for the i-th 
jackknife sample, $<r_{0}>$ is their average, and $N$ is the number of
jackknife samples.

In table~\ref{tbl:sigma_poisson_jackknife}, the jackknife errors 
$\sigma_{\mbox{\scriptsize Jackknife}}$ are compared with the Poisson errors
$\sigma_{\mbox{\scriptsize Poisson}}$ which were calculated without taking account of
the covariance.
The values of $\sigma_{\mbox{\scriptsize Jackknife}}$ are
systematically larger than those of $\sigma_{\mbox{\scriptsize Poisson}}$
by 0.3--1.0~$h^{-1}$Mpc.
It should be noted that the $\sigma_{\mbox{\scriptsize Jackknife}}$ 
includes a contribution not only from the covariance owing to the
galaxy clustering but also from
fluctuations of $r_{0}$ and $n_{bg}$ for each AGN field.
This was tested by performing a Monte Carlo simulation which simulated
the galaxy distribution according to the equation~(\ref{eq:model}) without
taking into account the covariance owing to the galaxy clustering.
According to the simple simulation, the difference of 0.3~$h^{-1}$Mpc for
the two errors can be explained by 50\% of flucatuions in projected 
cross-correlation
($\propto r_{0}^{\gamma}$) and the surface galaxy number density ($n_{bg}$).
The amount of the difference increases as the fluctuations increase.
Thus we can only measure the upper limit for the contribution of covariances,
which are less than 0.7~$h^{-1}$Mpc for the z4-B group, less than 
0.3~$h^{-1}$Mpc for the z1-D group, and so on.
The effect of the covariance is not so large that the results discussed 
above are not changed.

\section{Discussion}

We showed that the low redshift AGNs ($z$ = 0.3--0.9, z1 and z2) had 
clustering similar to that of a typical local galaxy.
This would imply that these AGNs are located in dark matter halos which have
a mass similar to that occupied by typical galaxies at the local Universe,
provided that there is a correlation between the clustering 
and the mass of the dark matter halo as suggested by many theoretical works
(e.g. \cite{Kaiser_1984, Mo_1996, Sheth_1999}).
%%
%% Coil et al (2007) reference
%%

No significant AGN luminosity dependence was observed for the z2 and z3 redshift
groups.
Since the luminosity of AGN is not constant and evolving with time, 
it may be possible for AGNs with different luminosities to have similar
environments and to be powered by the same mechanism at some epoch of their evolution.
The QSO lifetime, $t_{\mbox{\scriptsize Q}}$, which is a timescale for the most 
luminous phase of the activity, is expected to be in the range from 
10$^{6}$ -- 10$^{8}$ yr (\cite{Bell_2002,Martini_2004,Hopkins_2005}).
According to theoretical models (e.g. \cite{Sanders_1988}), AGNs are obscured 
by surrounding gas and dust
for a duration much longer than the period of the most luminous phase,
and thus they cannot be observed in the optical band and would be identified as
other types of galaxies such as ultra luminous infrared galaxies or sub-mm galaxies.
Therefore, no detection of the luminosity dependence of the clustering supports
the theoretical expectation of the short QSO lifetime (e.g. \cite{Adelberger_2005}).

We also showed that the bright AGNs at redshift 1.3--1.8 (z4-B) have relatively 
denser environments than typical local galaxies, and that the clustering strength
was much larger for bright galaxies ($M < -21$) than dim galaxies ($M \ge -21$).
The lower limit of the cross-correlation length of AGN and the bright galaxies
was 18~$h^{-1}$Mpc, which is extremely large compared to the local bright galaxies
measured by \citet{Zehavi_2005}.
Considering that our dataset is biased to blue star-forming galaxies
at higher redshift, we deduce that the clustering of blue star-forming
galaxies is enhanced around the high redshift AGNs compared to the local
Universe.

Under these circumstances, the chance of a major merger is expected to be
enhanced relative to that at low redshift, thus the major merger could be a
preferred mechanism to trigger the AGN activity at this redshift range.
This result indicates that there exists redshift dependence for the environment
of AGNs, that is, at higher redshift ($z > \sim$1.3) larger fraction of AGNs 
are produced in a more crowded region with blue star-forming galaxies, 
while at  lower redshift they are mostly produced in an environment which is 
common at the local Universe. 
AGNs at low redshift might be powered by different mechanism from high 
redshift AGNs, e.g. a secular process such as bars or other disk 
instabilities (\cite{Combes_2010} for a review).
It should be noted that it would be possible that these differences in fueling
is caused by differences in AGN luminosity since our high redshift sample is
biased to high luminosity.
To derive a more clear conclusion, we need to investigate a clustering
of low luminosity AGNs at high redshift.
The $\omega(r_{p})$ profile of the z4-B AGN group shows deviations 
from a power law function by 3.4$\sigma$, which shows a flat over 
density at $r_{p} < $2.5 Mpc.
This property could be attributed to the distribution of AGN location
in a dark matter halo; there may be a considerable number of AGNs
which have an offset of $\sim$2.5~Mpc from the center of galaxy 
distribution or a dark matter halo.
The $\omega(r_{p})$ for bright AGN groups have a tendency to show a deficit 
of galaxy at the smallest distance bin.
The significance of the deficit was calculated by fitting the model to data
excluding the data of the smallest distance bin and comparing data point 
with the extrapolation of the model function.
The significance of the deficit is 2.0, 0.4, and 3.1 $\sigma$ for z2-B,
z3-B and z4-B AGN groups, respectively.
This feature is not seen for any of the dim AGN groups.
This also might be explained by the distribution of AGN location inside the dark 
matter halo.
%%
%% Galaxy formation near the bright AGN may also be suppressed
%% due to the strong radiation from the AGN.
%%

%%
In this work, the error of $r_{0}$ is large compared to the experiments
measuring the redshift of all the target galaxies.
This is mostly due to the uncertainty of determination $\rho_{0}$, thus 
the accuracy of $\rho_{0}$ needs to be improved.
For that purpose, we need to estimate the accuracy of $\rho_{0}$ estimated
as described in section~\ref{sec:rho0} by comparing with 
spectroscopic observations for several fields.
Once the accuracy of the estimate of $\rho_{0}$ is examined experimentally, 
it would be possible for the uncertainty to be determined only
from uncertainty of the luminosity function.
Then we can reduce the uncertainty of $\rho_{0}$ by a factor two or more.
%%

%%
%% The current statistics is not enough for deriving definite conclusions.
%%
%% It is also necessary to consider the AGN properties at other wavelengths such
%% as X-ray, radio, and so on, since some of the AGNs may be hidden in dust or accretion
%% disk, and their activity could not be measured only from the optical luminosity.
%%
%% To ensure the validity of the relatively larger correlation length obtained for 
%% redshifts 0.8--1.8, we should raise the statistical significance by factor of ten or so, 
%% or it is necessary to remove the foreground and background galaxies by applying the
%% photometric redshift technique to enhance the signal to noise ratio.
%%
\section{Conclusions}

Using the Japanese Virtual Observatory, we were able to measure the clustering 
for 1,809 AGNs at redshifts 0.3--3.0 with optical 
brightness of $M_{V} = $ $-$30 to $-$20.
The AGN samples were divided into five redshift groups (z1--z5), and 
each redshift group was further divided into two brightness groups 
(e.g. z1-D, z2-D, z2-B ...).
We found evidence of clustering except for the highest redshift group (z5-B).
The cross-correlation lengths between the AGNs and galaxies were measured for each group.
We found that the clustering strength of the high redshift group (z4-B)
was larger than that of the low redshift group (z1-D), and it
was nearly consistent with that of QSO clustering at the same redshift.
We also showed that bright galaxies of absolute magnitude ($M_{AB} < -21$) were
more densely clustered around AGNs than the dimmer galaxies for the z4-B AGN group.
Considering that, at higher redshift ($z > 1$), the data used in this work are
based on optical observations and blue galaxies are expected to be preferentially 
detected in our dataset, we can deduce that blue star-forming galaxies are highly
clustered around the AGNs at these redshifts.
At the low redshift Universe the clustering of blue galaxies is smaller than 
that of red galaxies, and the deduced feature is different from that
seen at the local Universe.

The deviation of $\omega(r_{p})$ from the power law function was seen in the
result of the z4-B group at significance of 3.4$\sigma$.
There was an indication 
for suppression of cross-correlation at the smallest distance bin for the bright 
AGN groups, the significance was 2.0, 0.4, and 3.1~$\sigma$ for the z2-B, z3-B, and
z4-B group, respectively.
These features could be related to the distribution of AGN location in
a dark matter halo, and be related to fueling mechanism of the AGN.
If a significant fraction of AGNs was located offset from the center of a dark 
matter halo or a galaxy distribution, the deviation from power law could be 
seen.
%% 
%% It is also possible that the galaxy formation near the AGN was
%% suppressed due to the strong radiation of the AGN.
%%

%%
We showed that the analysis method presented in this paper can be a powerful
probe for the small scale environment of the high redshift AGNs.
Our method can be used for tracing dim and blue star-forming galaxies at small
scale ($r<1$ Mpc), while the spectroscopic observations provide a precise 
information about large scale clustering for bright and red galaxies.
These two types of observations, therefore, are complementary, and
can provide important clues to understanding the origin of the AGN activity.
To derive more definite evidence for the AGN-galaxy clustering with this method,
more observational data are needed in view of the quantity and quality.
In addition, it should be noted that in this work the properties of galaxies 
against which the AGN are being correlated are different for each AGN group,
thus it is essential to understand the auto-correlation properties of 
the galaxies in each group.
Highly homogeneous observations of a huge number of AGN fields will be 
conducted with the Hyper Suprime-Cam, which will be installed on the 
Subaru Telescope around 2011.
This survey project will give more definite observational evidence on the 
evolution of an AGN.

\section*{Acknowledgments}
We thank an anonymous referee for 
referee for helpful comments that greatly improved the text.
This work was supported by the JSPS Core-to-Core
Program and Grant-in-aid for Information Science (15017289, 16016292, 18049074 and
19024070) carried out by the Ministry of Education, Culture, Sports, Science
and Technology (MEXT) of Japan.
YS is grateful for support under Grant-in-aid for Young Scientists (B) 
(17700085, 21740143) 
carried out by the MEXT of Japan.

\newpage

%%%%%%%%%%%%%%%%%%%%%%%%%%%%%%%%%%%%%%%

%% \input{ref-sorted.tex}

\clearpage

\begin{table*}
\caption{Distance ranges for background estimation in the calculation of $B_{QG}$}
\label{tab:distance_range_bg}
\begin{center}
\begin{tabular}{cccc} \hline
redshift group & redshift range & $r_{\mbox{\scriptsize bg,min}}$ & $r_{\mbox{\scriptsize bg,max}}$ \\
      &                & Mpc  & Mpc \\
\hline
   z1 & 0.3 -- 0.6     & 1.5  & 2.5 \\
   z2 & 0.6 -- 0.9     & 2.0  & 3.5 \\
   z3 & 0.9 -- 1.3     & 3.0  & 5.0 \\
   z4 & 1.3 -- 1.8     & 4.0  & 6.0 \\
   z5 & 1.8 -- 3.0     & 4.0  & 6.0 \\
\hline
\end{tabular}
\end{center}
\end{table*}

\begin{table*}
   \caption{Number of AGNs selected from SDSS, 2dF, other optical, X-ray and radio observations
for each redshift and AGN luminosity group.}
   \label{tbl:number_AGN}
   \begin{center}
   \begin{tabular}{crrrrrr} \hline
       group
     & SDSS      \footnotemark[$*$] 
     &  2dF      \footnotemark[$\dagger$] 
     & UV-OPT-IR \footnotemark[$\ddagger$] 
     & XRAY      \footnotemark[$\S$] 
     & RADIO     \footnotemark[$\|$] 
     & Total \\ \hline
z1-D & 478 (73.5\%)  &  51 (7.8\%)  & 96 (14.7\%) & 20 (3.1\%) & 6 (0.9\%)  & 651   \\ 
z2-D & 334 (79.5\%)  &  66 (15.7\%) & 13 (3.1\%)  &  7 (1.7\%) & 0 (0.0\%)  & 420   \\ 
z3-D &  68 (68.0\%)  &  20 (20.0\%) &  9 (9.0\%)  &  3 (3.0\%) & 0 (0.0\%)  & 100   \\ 
z2-B & 101 (89.4\%)  &   4 (3.5\%)  &  2 (1.8\%)  &  2 (1.8\%) & 4 (3.5\%)  & 113   \\ 
z3-B & 234 (85.4\%)  &  32 (11.7\%) &  4 (1.5\%)  &  2 (0.7\%) & 2 (0.7\%)  & 274   \\ 
z4-B & 107 (75.4\%)  &  21 (14.8\%) & 10 (7.0\%)  &  4 (2.8\%) & 0 (0.0\%)  & 142   \\ 
z5-B &  76 (69.7\%)  &  21 (19.3\%) &  8 (7.3\%)  &  4 (3.7\%) & 0 (0.0\%)  & 109   \\  
total& 1398 (77.3\%)  & 215 (11.9\%) & 142 (7.8\%)  & 42 (2.3\%) & 12 (0.7\%) & 1809 \\
\hline
\multicolumn{3}{@{}l@{}}{
   \hbox to 0pt{\parbox{160mm}{\footnotesize
\par\noindent
\footnotemark[$*$] Number of AGN samples contained in the SDSS DR5 QSO catalog (4th Ed.) 
 (\cite{Schneider_2007}). This number includes samples contained also in the 2dF QSO catalog.
\par\noindent
\footnotemark[$\dagger$] Number of AGN samples contained in the 2dF QSO catalog
 (\cite{Croom_2004}) and not contained in the SDSS QSO catalog.
\par\noindent
\footnotemark[$\ddagger$] Number of AGN samples selected based on the optical properties.
References of each AGN are:
\cite{Zakamska_2003}, % 2664
\cite{Reyes_2008}, % 1889
\cite{Abazajian_2004}, % 3
\cite{Collinge_2005}, % 462
\cite{Schneider_2005}, % 2067
\cite{Prescott_2006}, % 1832
\cite{Barton_2006}, % 163
\cite{Zhan_1989}, % 2676
\cite{Williams_2002}, % 2542
\cite{Richards_2001}, % 1893
\cite{Hall_2000}, % 935
\cite{Veron_1990}, % 2425
\cite{Rowan-Robinson_1990}, % 1930
\cite{McIntosh_2004}, % 1540
%% \cite{Margon_1983}, % 1446
\cite{Liu_1999}, % 1369
\cite{Leipski_2005}, % 1345
\cite{LaFranca_1992}, % 1301
%%\cite{Koo_1986}, % 1256
\cite{Kniazev_2004}, % 1232
\cite{Hewett_1995}, % 1012
%%\cite{Gaston_1983}, % 799
\cite{Cristiani_1995}, % 533
\cite{Boyle_1990}, % 283
\cite{Hennawi_2006}, % 1001
\cite{Hall_1996}, % 934
%%\cite{Croom_2009}, % 540
%%\cite{Cohen_1999}, % 448
%%\cite{Boyle_1991}, % 284
%%\cite{Boller_1989}, % 237
%%\cite{Adelman-McCarthy_2006}, % 13
\cite{Zhan_1987}, % 2673
\cite{Schneider_1994}, % 2058
\cite{Hewett_1991}, % 1010
\cite{Sharp_2002}, % 2092
%%\cite{Wolf_1999}, % 2603
%%\cite{Schneider_1999}, % 2061
%%\cite{Schade_1996}, % 2028
%%\cite{Kennefick_1997}, % 1206
\cite{Abazajian_2009}, % 4
\cite{Hoag_1982}, % 1040
%%\cite{Weedman_1985}, % 2488
\cite{Usher_1983}, % 2371
\cite{Monk_1988}, % 1605
\cite{He_1984}, % 986
\cite{Dressler_1992}, % 649
%%\cite{Borra_1996} % 272
\cite{Cristiani_1989} % 529  1989A&AS...77..161C
\cite{Zhou_2006} % 2689 2006ApJS..166..128Z
\cite{Schmidt_1986} % 2042 1986ApJ...306..411S
%%\cite{Trevese_1994} % 2324 1994ApJ...433..494T
%%\cite{VandenBerk_2005} % 2380 2005AJ....129.2047V

%%
\par\noindent
\footnotemark[$\S$] Number of AGN samples selected based on X-ray observations.
References of each AGN are:
\cite{Silverman_2005}, % 2114
\cite{Steffen_2004}, % 2184
\cite{Lehmann_2000}, % 1342
\cite{Watanabe_2002}, % 2481
%%\cite{Stocke_1991}, % 2242
%%\cite{Stern_2002}, % 2216
%%\cite{Schreier_2001}, % 2074
\cite{Schmidt_1998}, % 2047
\cite{Molthagen_1997}, % 1598
%%\cite{Ohta_2003}, % 1682
\cite{Lehmann_2001}, % 1343
\cite{LaFranca_2002}, % 1304
\cite{Hasinger_2002}, % 957
\cite{Grindlay_1993}, % 900
\cite{Gioia_2003}, % 825
%%\cite{Fischer_1998}, % 752
\cite{Eckart_2006}, % 664
\cite{Dewangan_2001}, % 597
\cite{Crawford_2002}, % 521
\cite{Caccianiga_2004}, % 370
\cite{Branduardi-Raymont_1985}, % 298
\cite{Bauer_2000}, % 170
%%\cite{Basilakos_2002}, % 166
\cite{Barger_2002}, % 152
%%\cite{Barger_2001}, % 151
%%\cite{Akiyama_2003}, % 27
%%\cite{Szokoly_2004}, % 2272
\cite{McHardy_1998}, % 1538
\cite{Miller_2002}, % 1572
%%\cite{Hornschemeier_2001}, % 1054
\cite{Georgantopoulos_2004}, % 813
\cite{Fiore_2003}, % 751
\cite{Bower_1996}, % 281
%%\cite{Barger_2003}, % 153
\cite{Bade_1995}, % 120
%%\cite{Trevese_2007}, % 2325
%%\cite{Treister_2009}, % 2323
%%\cite{Dewangan_2002}, % 598
\cite{Mason_2000}, % 1513
%%\cite{Boyle_1997}, % 287
\cite{Apr_2001}, % 106
\cite{Crampton_1997}, % 520
%%\cite{Green_2004} % 866

%%

%%
\par\noindent
\footnotemark[$\|$] Number of AGN samples selected from Radio observations.
References of each AGN are:
\cite{Becker_2001}, % 179
%%\cite{Francis_2000}, % 775
\cite{Glikman_2004}, % 833
\cite{Healey_2008}, % 988
\cite{Hes_1996}, % 1009
\cite{Lynds_1966}, % 1395
\cite{Marziani_1996}, % 1494
\cite{Rowan-Robinson_2004}, % 1930
%%\cite{Tran_1998}, % 2319
\cite{White_2000}, % 2528
%%\cite{Wiloson_1980}, % 2537
\cite{Wold_2000} % 2602
\cite{Buchalter_1998} % 332  1998ApJ...494..503B
   }\hss}}
   \end{tabular}
   \end{center}
\end{table*}

\begin{table*}
  \caption{Number of AGN samples for each AGN group and observation band.}
  \label{tbl:number_band}
  \begin{center}
  \begin{tabular}{crrrrrrr}
\hline
group & B & V & R & I & i' & z' & K \\
\hline
z1-D &   13 &  7 & 21 &  9 & 13 &  4 & 584 \\
z2-D &    0 &  6 & 18 & 16 &  4 &  2 & 374 \\
z3-D &    2 &  3 &  4 &  6 &  5 &  4 &  76 \\
z2-B &    1 &  1 &  9 &  7 &  3 &  1 &  91 \\
z3-B &   10 & 15 & 25 & 19 & 11 &  8 & 186 \\
z4-B &   11 & 15 & 45 & 32 & 21 &  9 &   9 \\
z5-B &   12 & 19 & 38 & 18 & 12 &  5 &   5 \\
\hline
  \end{tabular}
  \end{center}
\end{table*}

\begin{table*}
  \caption{Fitting parameters of projected cross correlation analysis}
  \label{tbl:fit_param}
  \begin{center}
  \begin{tabular}{lcccccrccc}
\hline
group &
$z$        & 
$<z>$\footnotemark[$*$]  &  
M$_{V}$\footnotemark[$\dagger$]  & 
$n_{\mbox{\scriptsize AGN}}$\footnotemark[$\ddagger$] & 
r$_{0}$\footnotemark[$\|$]       & 
$<n_{bg}>$\footnotemark[$\S$]      & 
$<\rho_{0}>$\footnotemark[$\#$]  \\
%
%  *, \dagger, \ddagger, \S, \|, \#,
%
             &  
             &  
             &
             & 
mag          &
h$^{-1}$Mpc   &
Mpc$^{-2}$ & 
$10^{-3}$Mpc$^{-3}$ \\
\hline
z1-D & 0.3 -- 0.6 & 0.46 & $-$25.0 -- $-$20.0 & 651 &  4.7$^{+1.3}_{-0.7}$ & 22.1 $\pm$ 0.07 & 4.3$^{+0.7}_{-1.1}$ \\
z2-D & 0.6 -- 0.9 & 0.73 & $-$25.0 -- $-$20.0 & 420 &  5.8$^{+1.9}_{-1.1}$ & 10.1 $\pm$ 0.05 & 1.9$^{+0.4}_{-0.6}$ \\
z3-D & 0.9 -- 1.3 & 1.01 & $-$25.0 -- $-$20.0 & 100 &  7.6$^{+3.2}_{-1.9}$ & 11.2 $\pm$ 0.07 & 1.6$^{+0.4}_{-0.5}$ \\
z2-B & 0.6 -- 0.9 & 0.79 & $-$30.0 -- $-$25.0 & 113 &  6.3$^{+2.4}_{-1.5}$ & 13.9 $\pm$ 0.10 & 2.3$^{+0.4}_{-0.6}$\\
z3-B & 0.9 -- 1.3 & 1.06 & $-$30.0 -- $-$25.0 & 274 &  5.1$^{+2.7}_{-1.5}$ & 11.1 $\pm$ 0.04 & 1.4$^{+0.4}_{-0.5}$  \\
z4-B & 1.3 -- 1.8 & 1.54 & $-$30.0 -- $-$25.0 & 142 & 11.1$^{+6.1}_{-2.7}$ & 19.4 $\pm$ 0.07 & 1.3$^{+0.4}_{-0.6}$ \\
z5-B & 1.8 -- 3.0 & 2.07 & $-$30.0 -- $-$25.0 & 109 & $<$ 13.1             & 14.4 $\pm$ 0.07 & 0.49$^{+0.19}_{-0.24}$ \\
\hline
z1-D (IR)  & 0.3 -- 0.6 & 0.46 & $-$25.0 -- $-$20.0 & 584 & 6.8$^{+1.9}_{-1.0}$ &  7.73$\pm$0.05 & 2.2$^{+0.4}_{-0.6}$ \\
z1-D (OPT) & 0.3 -- 0.6 & 0.46 & $-$25.0 -- $-$20.0 & 67  & 3.0$^{+1.5}_{-1.1}$ & 147.0$\pm$0.67 & 22.3$^{+2.8}_{-4.6}$  \\
z2-D (IR)  & 0.6 -- 0.9 & 0.73 & $-$25.0 -- $-$20.0 & 374 & 7.4$^{+4.7}_{-2.1}$ &  3.50$\pm$0.03 & 0.56$^{+0.22}_{-0.27}$ \\
z2-D (OPT) & 0.6 -- 0.9 & 0.74 & $-$25.0 -- $-$20.0 & 46  & 5.0$^{+1.6}_{-1.0}$ &  63.5$\pm$0.34 & 12.7$^{+1.8}_{-2.9}$  \\
\hline
\multicolumn{3}{@{}l@{}}{
   \hbox to 0pt{\parbox{160mm}{\footnotesize
\par\noindent
\footnotemark[$*$] average redshift
\par\noindent
\footnotemark[$\dagger$]  K-corrected absolute magnitude range
\par\noindent
\footnotemark[$\ddagger$] number of AGNs
\par\noindent
\footnotemark[$\|$] correlation length, the quoted error contains systematic
error due to uncertainty of $\rho_{0}$ and one sigma statistical error.
The error does not include covariance of number densities at each distance bins.
\par\noindent
\footnotemark[$\S$] projected galaxy number density for background
\par\noindent
\footnotemark[$\#$] average galaxy number density at the AGN redshift
   }\hss}}
  \end{tabular}
  \end{center}
\end{table*}

\begin{table*}
  \caption{Comparison between two error estimates for $r_{0}$}
  \label{tbl:sigma_poisson_jackknife}
  \begin{center}
  \begin{tabular}{ccc}
\hline
group & 
$\sigma_{\mbox{\scriptsize Poisson}}$\footnotemark[$*$] &
$\sigma_{\mbox{\scriptsize Jackknife}}$\footnotemark[$\dagger$] \\
\hline
z1-D   & 0.4 & 0.7 \\
z2-D   & 0.6 & 1.0 \\
z3-D   & 1.2 & 2.1 \\
z2-B   & 1.0 & 2.0 \\
z3-B   & 1.0 & 1.9 \\
z4-B   & 1.2 & 1.9 \\
\hline
\multicolumn{3}{@{}l@{}}{
   \hbox to 0pt{\parbox{160mm}{\footnotesize
\par\noindent
\footnotemark[$*$] error of $r_{0}$ estimated by using Poisson errors of
number density.
\par\noindent
\footnotemark[$\dagger$]  error of $r_{0}$ estimated by the Jackknife method.
   }\hss}}
  \end{tabular}
  \end{center}
\end{table*}

\clearpage

\clearpage

\end{document}